\documentclass[twocolumn,showpacs,preprintnumbers,amsmath,nofootinbib,amssymb]{revtex4-1}

\usepackage{graphicx}
\usepackage{dcolumn}
\usepackage{bm}
\usepackage{amsfonts}
\usepackage{mathrsfs}

\def\bequ{\begin{equation}}
\def\eequ{\end{equation}}
\def\be{\begin{equation}}
\def\ee{\end{equation}}

\begin{document}

\title{Maxwell perturbations on Kerr-anti-de Sitter: \\
quasinormal modes,
 superradiant instabilities and vector clouds}

\author{Mengjie Wang}
\email{mengjie.wang@ua.pt}
\author{Carlos Herdeiro}
\email{herdeiro@ua.pt}
\affiliation{\vspace{2mm}Departamento de F\'\i sica da Universidade de Aveiro and CIDMA \\
Campus de Santiago, 3810-183 Aveiro, Portugal \vspace{1mm}}%

\date{December 2015}

\begin{abstract}
Scalar and gravitational perturbations on Kerr-anti-de Sitter (Kerr-AdS) black holes have been addressed in the literature and have been shown to exhibit a rich phenomenology. In this paper we complete the analysis of bosonic fields on this background by studying Maxwell perturbations, focusing on superradiant instabilities and vector clouds. For this purpose, we solve the Teukolsky equations numerically, imposing the boundary conditions we have proposed in~\cite{Wang:2015goa} for the radial Teukolsky equation. As found therein, two Robin boundary conditions  can be imposed for Maxwell fields on Kerr-AdS black holes, one of which produces a new set of quasinormal modes even for Schwarzschild-AdS black holes. Here, we show these different boundary conditions produce two different sets of superradiant modes. Interestingly the ``new modes'' may be unstable in a larger parameter space. We then study stationary Maxwell clouds, that exist at the threshold of the superradiant instability, with the two Robin boundary conditions. These clouds, obtained at the linear level, indicate the existence of a new family of  black hole solutions at the nonlinear level, within the Einstein-Maxwell-AdS system, branching off from the Kerr-Newman-AdS family. As a comparison with the Maxwell clouds, scalar clouds on Kerr-AdS black holes are also studied, and it is shown there are Kerr-AdS black holes that are stable against scalar, but not vector modes, with the same ``quantum numbers".
\end{abstract}

\maketitle

\section{Introduction}

The global structure of asymptotically anti-de Sitter (AdS) spacetimes allows interesting novel features, as compared to asymptotically flat spacetimes. For instance, when a rotating black hole (BH) exists in the bulk of an asymptotically AdS spacetime, superradiant instabilities (see the recent review~\cite{Brito:2015oca}) can be triggered by a massless field, in contrast to the asymptotically flat case wherein such instabilities only arise for massive fields. Superradiant instabilities occur for a bosonic field wave, with the time/azimuthal dependence $e^{-i\omega t+im\phi}$, impinging on a Kerr-AdS BH with angular velocity $\Omega_H$,  when the condition Re$(\omega) < m \Omega_H$ is satisfied, yielding an amplified scattered wave. The amplified wave can be reflected at the AdS boundary, and the wave bounces back and forth between the BH and the AdS boundary, leading to the instability~\cite{Cardoso:2004hs,Cardoso:2006wa,Dias:2011at,Uchikata:2009zz,Cardoso:2013pza,Wang:2014eha}.

Superradiant instabilities of Kerr-AdS BHs triggered by both scalar~\cite{Uchikata:2009zz} and gravitational fields~\cite{Cardoso:2013pza} have already been discussed. Here we shall consider Maxwell perturbations of the Kerr-AdS background, which remained, hitherto, unaddressed\footnote{For recent studies of Maxwell perturbations on pure AdS see, $e.g.$,~\cite{Herdeiro:2015vaa,Holzegel:2015swa}.}. Central to this analysis are the boundary conditions to be imposed. Early perturbation studies of spin fields on Schwarzschild-AdS BHs imposed field vanishing boundary conditions, see $e.g.$~\cite{Cardoso:2001bb}, within the Regge-Wheeler formalism. In Kerr-AdS BHs, on the other hand, perturbation equations for spin fields can only be separated and decoupled in the Teukolsky formalism; then imposing boundary conditions becomes a trickier problem. Even in a simpler Schwarzschild-AdS BH, it is not clear, in general, how to produce the same results both in the Regge-Wheeler formalism and in the Teukolsky formalism.

A new perspective on the issue of boundary conditions in asymptotically AdS spacetimes was recently put forward~\cite{Wang:2015goa}. We have proposed the following simple principle: the physical boundary conditions that should be imposed on perturbations of asymptotically AdS spacetimes should require the energy flux to vanish at the AdS boundary. As shown in an Appendix herein, this requirement also implies the vanishing of the angular momentum flux at the AdS boundary. Applying this principle to Maxwell perturbations on Schwarzschild-AdS BHs, we have observed in~\cite{Wang:2015goa} the existence of two sets of quasinormal modes, one of which had not been discussed in the literature. As a sequel of~\cite{Wang:2015goa}, here we shall  apply the two sets of boundary conditions obtained therein to Maxwell perturbation on Kerr-AdS BHs. The inclusion of rotation leads to new physical phenomena that we shall explore.

The first goal of this paper is to address superradiant instabilities triggered by the Maxwell field on Kerr-AdS BHs. To achieve it, we use both an analytical matching scheme as well as a numerical method to explore the problem. The former method provides an intuitive way to understand how these two boundary conditions produce different instabilities, in the small BH and slow rotation regime. The latter method provides a technique to understand the problem in a larger region of the parameter space. We find that Maxwell fields can trigger stronger instabilities than scalar fields~\cite{Uchikata:2009zz} in the superradiant regime, for both boundary conditions.

The second goal of this paper is to study vector \mbox{clouds} on Kerr-AdS BHs. Stationary clouds~\cite{Hod:2012px} are bound \mbox{state} solutions of test fields on a rotating background, at linear level. They exist at the threshold of the superradiant instabilities triggered by that test field. Recently, a considerable number of studies of such clouds has \mbox{appeared} in the literature, mostly in asymptotically flat spacetimes~\cite{Hod:2012px,Hod:2013zza,Shlapentokh-Rothman:2013ysa,Herdeiro:2014goa,Hod:2014baa,Benone:2014ssa,Hod:2014sha,Hod:2014pza,Herdeiro:2014pka,Benone:2014nla,Hod:2015ota,Wilson-Gerow:2015esa,Benone:2015jda,Herdeiro:2015gia,Hod:2015ynd}, but also in asymptotically AdS \mbox{spacetimes}~\cite{Dias:2011at,Cardoso:2013pza}\footnote{Analogous clouds around charged BHs in a cavity have been addressed in~\cite{Sakalli:2015uka,Li:2015bfa}.}. Most of these studies have addressed scalar field clouds\footnote{Superradiance onset curves for gravitational perturbations on Kerr-AdS, which --  as the existence lines for stationary clouds -- identify the backgrounds supporting the zero-mode of the perturbation,  have been studied in~\cite{Cardoso:2013pza}, and the corresponding ``hairy'' BH solutions have been constructed in~\cite{Dias:2015rxy}.} (even though \textit{marginal} clouds~\cite{Degollado:2013eqa} have been considered for a Proca field~\cite{Sampaio:2014swa} in a charged BH background).  Here we perform a study of Maxwell clouds, which can exist around rotating BHs in asymptotically AdS spacetimes, since the AdS global structure dispenses with the mass term. As a comparison with the Maxwell clouds on Kerr-AdS BHs we also consider scalar clouds in the same background.

It was proposed in~\cite{Herdeiro:2014goa,Herdeiro:2014ima} that the existence of stationary clouds of a given test field, as a \textit{zero-mode} of the superradiant instability, indicates the existence of new families of ``hairy'' BH solutions, at fully nonlinear level, such as the Kerr BHs with scalar hair found (numerically) in ~\cite{Herdeiro:2014goa}, whose existence was recently formally proved~\cite{Chodosh:2015oma}. It is an open issue if these hairy BHs may be formed dynamically, as the end point of the instability. Interesting evidence in this direction was reported recently~\cite{Dolan:2015dha} for the case of Reissner-Nordstr\"om BHs in a cavity, following the earlier discussion of superradiant instabilities in this setup~\cite{Herdeiro:2013pia,Hod:2013fvl,Degollado:2013bha}. The existence of Maxwell clouds on Kerr-AdS BHs has, therefore, the interesting implication that new families of solutions of charged rotating BHs exist, within the Einstein-Maxwell-AdS system, besides the well known Kerr-Newman-AdS family, and branching off from the latter.

This paper is organized as follows. In Section~\ref{seceq}, the Kerr-AdS background geometry is described, the Teukolsky equations for the Maxwell field are presented and the corresponding boundary conditions are illustrated. In Section~\ref{secanalytical}, the analytical matching method is developed to study quasinormal modes in the small BH and slow rotation regime. In Section~\ref{secnumerical}, the numerical method is briefly introduced and the numerical results are presented, including a brief discussion on scalar perturbations, as a comparison with the vector case. Final remarks and conclusions are presented in the last section. As an Appendix, we provide a demonstration that the vanishing energy flux boundary conditions imply also that the angular momentum flux  vanishes.

\section{Maxwell equations on Kerr-AdS BHs and the boundary conditions}
\label{seceq}
In this section we briefly review the properties of Kerr-AdS BHs, Teukolsky equations of the Maxwell fields and the corresponding boundary conditions, to introduce the fundamental quantities and set the notations.
\subsection{Kerr-AdS BHs}
The line element for a Kerr-AdS BH, in Boyer-Lindquist coordinates, can be written as
\begin{eqnarray}
ds^2&=&\dfrac{\Delta_r}{\rho^2\Xi^2}\Big(dt-a\sin^2\theta d\varphi\Big)^2-\rho^2\left(\dfrac{dr^2}{\Delta_r}+\dfrac{d\theta^2}{\Delta_\theta}\right)\nonumber\\&-&\dfrac{\Delta_\theta \sin^2\theta}{\rho^2\Xi^2}\Big(adt-(r^2+a^2)d\varphi\Big)^2\;,\label{metric}
\end{eqnarray}
with metric functions
\begin{eqnarray}
\rho^2&=&r^2+a^2\cos^2\theta\;,\;\;\Delta_r=(r^2+a^2)\left(1+\frac{r^2}{L^2}\right)-2Mr\;,\nonumber\\
\Delta_\theta&=&1-\dfrac{a^2\cos^2\theta}{L^2}\;,\;\;\Xi=1-\dfrac{a^2}{L^2}\;,\label{metricfunc}
\end{eqnarray}
where $L$ is the AdS radius and $M$, $a$ stand for the mass and spin parameters, related to the BH energy and angular momentum.
In this frame, the angular velocity of the event horizon and the Hawking temperature are given by
\begin{eqnarray}
\Omega_H&=&\dfrac{a}{r_+^2+a^2}\;,\label{Horvel}\\
T_H&=&\dfrac{1}{\Xi}\left[\dfrac{r_+}{2\pi}\left(1+\dfrac{r_+^2}{L^2}\right)\dfrac{1}{r_+^2+a^2}-\dfrac{1}{4\pi r_+}\left(1-\dfrac{r_+^2}{L^2}\right)\right]\;,\nonumber\\\label{Temp}
\end{eqnarray}
where the event horizon $r_+$ is determined as the largest root of $\Delta_r(r_+)=0$. For a given $r_+$, the mass parameter $M$ can be expressed as
\begin{equation}
M=\dfrac{(r_+^2+a^2)(L^2+r_+^2)}{2r_+L^2}\;.\nonumber
\end{equation}
The rotation parameter $a$ satisfies the following constraints
\begin{eqnarray}
&&\dfrac{a}{L}\leq \dfrac{r_+}{L} \sqrt{\dfrac{3r_+^2+L^2}{L^2-r_+^2}}\;,\;\;\;\;\;\text{for}\;\; \dfrac{r_+}{L}<\dfrac{1}{\sqrt{3}}\;,\label{a1}\\
&&\dfrac{a}{L}<1\;,\;\;\;\;\;\;\;\;\;\;\;\;\;\;\;\;\;\;\;\;\;\;\;\;\;\;\;\text{for}\;\; \dfrac{r_+}{L}\geq\dfrac{1}{\sqrt{3}}\;,\label{a2}
\end{eqnarray}
by requiring that a horizon exists and to avoid singularities. Note that the equality condition in Eq.~\eqref{a1} corresponds to extremal BHs.
\subsection{Teukolsky equations of the Maxwell field}
Studies on linear perturbation of  arbitrary spin massless fields on Kerr-dS BHs can be traced back to the early 1980s~\cite{Khanal:1983vb} (see also~\cite{Wu:2003qc,Yoshida:2010zzb}); recently the analogous equation was derived for a Kerr-AdS BH~\cite{Dias:2012pp}, using a different coordinate system and in a different context.

In the Newman-Penrose formalism, the Maxwell equations are described in terms of three complex scalars, two of which are independent. These two scalars are denoted by $\phi_0$ and $\phi_2$, and can be expanded as
\begin{eqnarray}
\phi_0&=&e^{-i\omega t+im\varphi} R_{+1}(r)S_{+1}(\theta)\;,\nonumber\\
\phi_2&=&\dfrac{B}{2(\bar{\rho}^\ast)^2}e^{-i\omega t+im\varphi} R_{-1}(r)S_{-1}(\theta)\;,\label{fielddeccom}
\end{eqnarray}
where the relative amplitude between $\phi_0$ and $\phi_2$ is set by $B$, which is a positive root of~\cite{Wu:2003qc}
\begin{equation}
B^2=\lambda^2-4\Xi^2\omega(\omega a^2-ma)\;,\label{Beq}
\end{equation}
and where $\lambda$ is a separation constant.

In the following, we present both the radial and angular equations governing a spin $s (s=\pm1)$ perturbation, adapted to our frame and notation.
The radial equation is
\begin{equation}
\Delta_r^{-s}\dfrac{d}{dr}\left(\Delta_r^{s+1}\dfrac{d R_{s}(r)}{dr}\right)+H(r)R_{s}(r)=0\;,\label{radialeq}
\end{equation}
with
\begin{eqnarray}
H(r)=\dfrac{K_r^2-i s K_r \Delta_r^\prime}{\Delta_r}+2isK_r^\prime+\dfrac{s+|s|}{2}\Delta_r^{\prime\prime}
+\dfrac{a^2}{L^2}-\lambda\;,\nonumber
\end{eqnarray}
where
\begin{eqnarray}
K_r=[\omega(r^2+a^2)-am]\Xi\;.\label{Kreq}
\end{eqnarray}
The angular equation, on the other hand, is
\begin{equation}
\dfrac{d}{du}\left(\Delta_u\dfrac{dS_{lm}}{du}\right)+A(u)S_{lm}=0\;,\label{angulareq}
\end{equation}
with $u=\cos\theta$, and
\begin{equation}
A(u)=-\dfrac{K_u^2}{\Delta_u}-4smu\dfrac{\Xi}{1-u^2}+\lambda-|s|-2(1-u^2)\dfrac{a^2}{L^2}\;,\nonumber
\end{equation}
where
\begin{eqnarray}
K_u&=&\Big(\omega a (1-u^2)+(s u-m)\Big)\Xi\;,\nonumber\\
\Delta_u&=&(1-u^2)\left(1-\dfrac{a^2}{L^2}u^2\right)\;.\nonumber
\end{eqnarray}
As we addressed in~\cite{Wang:2015goa}, the Teukolsky equations for $s=+1$ and $s=-1$ encode the same information. For concreteness and without loss of generality, we specify $s=-1$ in the following and consider the corresponding BCs, which are illustrated in the next subsection.

\subsection{Boundary conditions}
To study quasinormal modes, superradiant modes and vector clouds, we have to assign physically relevant boundary conditions to the Maxwell perturbations. Since the radial equation~\eqref{radialeq} and the angular equation~\eqref{angulareq} are coupled with each other through $\lambda$ and $\omega$, we have to impose boundary conditions for both equations. We address the boundary conditions for the radial equation firstly. At the horizon, ingoing boundary conditions should be imposed
\begin{eqnarray}
R_{-1}\sim(r-r_+)^\rho\;,
\end{eqnarray}
with
\begin{equation}
\rho=1-\dfrac{i(\omega-m\Omega_H)}{4\pi T_H}\;.\nonumber
\end{equation}
At infinity, the asymptotic analysis of the radial Teukolsky equation~\eqref{radialeq} with $s=-1$ yields the solution
\begin{equation}
R_{-1} \sim \;\alpha^{-} r+\beta^{-}+\mathcal{O}(r^{-1})\;,\label{asysol}
\end{equation}
where $\alpha^{-}$ and $\beta^{-}$ are two integration constants. Taking the viewpoint that the AdS boundary may be regarded as a perfectly reflecting mirror, we ask the energy flux to vanish asymptotically. This requirement leads to the following two Robin boundary conditions~\cite{Wang:2015goa}
\begin{eqnarray}
\dfrac{\alpha^{-}}{\beta^{-}}&=&\dfrac{2i\omega\Xi}{B-\lambda+2\omega^2\Xi^2L^2}\;,\label{bc1}\\
\dfrac{\alpha^{-}}{\beta^{-}}&=&\dfrac{2i\omega\Xi}{- B-\lambda+2\omega^2\Xi^2L^2}\;,\label{bc2}
\end{eqnarray}
where $B$ is given in Eq.~\eqref{Beq}. One may ask if the angular momentum flux also vanishes with the above two Robin boundary conditions. Physically this is to be expected, as an angular momentum flux without energy flux would violate the dominant energy condition and that is not expected to happen for the Maxwell field. In Appendix~\ref{AMF} we will prove that the angular momentum flux also vanishes, indeed. For more details on these Robin boundary conditions, we refer readers to~\cite{Wang:2015goa}.

To solve the angular equation~\eqref{angulareq}, we shall require its solutions to be regular at the singular points $\theta=0$ and $\theta=\pi$. This determines uniquely the set of angular functions labelled by $\ell$ and $m$.

\section{Analytic matching calculations}
\label{secanalytical}
In this section, we present an analytic calculation of quasinormal frequencies for a Maxwell field on a Kerr-AdS BH, with the two Robin boundary conditions discussed in the previous section. Such calculations can be used to illustrate how these Robin boundary conditions generate unstable modes.

Making use of the standard matching procedure, we shall first divide the space outside the event horizon into two regions: the \textit{near region}, defined by the condition $r-r_+\ll1/\omega$, and the \textit{far region}, defined by the condition $r_+\ll r-r_+$. Then, we further require the condition $r_+\ll1/\omega$, so that an overlapping region exists where solutions obtained in the near region and in the far region are both valid. In the following analysis we focus on small AdS BHs $(r_+\ll L)$ with slow rotation $(a\ll r_+)$. The former condition allows treating the frequencies for the Kerr-AdS BH as a perturbation of the AdS normal frequencies; the latter condition, together with $\omega r_+\ll1$, implying $\omega a\ll 1$ and $a\ll L$, allows approximating the angular equation for the spin-weighted AdS-spheroidal harmonics by the spin-weighted spherical harmonics, so that the separation constant becomes
\begin{equation}
 \lambda\simeq\ell(\ell+1)\;,\;\;\;\text{with}\;\;\ell=1,2,3\;,\;\cdot\cdot\cdot\;,\label{eigenvalues}
\end{equation}
where $\ell$ is the angular quantum number.

\subsection{Near region solution}
In the near region, under the small BH, $r_+\ll L$, and the slow rotation, $a\ll r_+$, approximations, Eq.~\eqref{radialeq} becomes
\begin{eqnarray}
\Delta_r R_{-1}''+\left(\dfrac{(r_+-r_-)^2\hat{\omega}}{\Delta_r}-\lambda\right)R_{-1}=0 \;,\label{neareq1}
\end{eqnarray}
with
\begin{equation}
\hat{\omega}=\left(\bar{\omega}+\dfrac{i}{2}\right)^2+\dfrac{1}{4}\;,\;\;\;\bar{\omega}=\left(\omega-m\Omega_H\right)\Xi\dfrac{r_+^2+a^2}{r_+-r_-}\;,\nonumber
\end{equation}
where $\Omega_H$ is the angular velocity of the event horizon, given by Eq.~\eqref{Horvel}.
It is convenient to define a new dimensionless variable
\begin{equation}
z\equiv \dfrac{r-r_+}{r-r_-}\;,\nonumber
\end{equation}
to transform Eq.~\eqref{neareq1} into
\begin{equation}
z(1-z)\frac{d^2R_{-1}}{dz^2}-2z\frac{dR_{-1}}{dz}+\left(\dfrac{\hat{\omega}(1-z)}{z}-\dfrac{\lambda}{1-z}\right)R_{-1}=0\;.\label{neareq2}
\end{equation}
The above equation can be solved in terms of the hypergeometric function
\begin{equation}
R_{-1}\sim z^{1-i\bar{\omega}}(1-z)^\ell\;F(\ell+1,\ell+2-2i\bar{\omega},2-2i\bar{\omega};z)\;,\label{nearsol}
\end{equation}
where an ingoing boundary condition has been imposed.

The near region solution, Eq.~\eqref{nearsol}, must be expanded for large $r$, in order to perform the matching with the far region solution below. To achieve this we take the $z\rightarrow1$ limit, and obtain
\begin{eqnarray}
R_{-1} \;\sim \;&& \Gamma(2-2i\bar{\omega})\left[\dfrac{R^{\rm near}_{-1,1/r}}{r^{\ell}} +R^{\rm near}_{-1,r} r^{\ell+1}\right]
\;,\label{nearsolfar}
\end{eqnarray}
where
\begin{eqnarray}
R^{\rm near}_{-1,1/r} & \equiv & \dfrac{\Gamma(-2\ell-1) (r_+ - r_-)^\ell}{\Gamma(-\ell)\Gamma(1-\ell-2i\bar{\omega})} \ ,\nonumber \\
R^{\rm near}_{-1,r} &\equiv& \dfrac{\Gamma(2\ell+1)(r_+-r_-)^{-\ell-1}}{\Gamma(\ell+1)\Gamma(\ell+2-2i\bar{\omega})} \ ,
\end{eqnarray}
by using the properties of the hypergeometric function~\cite{abramowitz+stegun}.

%
\subsection{Far region solution}
In the far region, $r-r_+\gg r_+$, the BH effects can be neglected ($M\rightarrow0, a\rightarrow0$) so that
\begin{equation}
\Delta_r\simeq r^2 \left(1+\dfrac{r^2}{L^2}\right)\; .\nonumber
\end{equation}
Then Eq.~\eqref{radialeq} becomes
\begin{eqnarray}
&&\Delta_rR_{-1}''(r)+\left(\dfrac{K_r^2+i K_r \Delta_r^\prime}{\Delta_r}-2iK_r^\prime
-\ell(\ell+1)\right)R_{-1}(r)\nonumber\\
&&=0\;,\label{fareq1}
\end{eqnarray}
with $K_r=\omega r^2$.

The general solution for Eq.~\eqref{fareq1} is
\begin{eqnarray}
&&R_{-1}=\;r^{\ell+1}(r-iL)^{\frac{\omega L}{2}}(r+iL)^{-\ell-\frac{\omega L}{2}}\Big[C_1F\Big(\ell,\ell+1\Big.\Big.\nonumber \\ && \Big.\Big.+\omega L,2\ell+2;\dfrac{2r}{r+iL}\Big)-2^{-2\ell-1}C_2\left(1+\dfrac{iL}{r}\right)^{2\ell+1}\Big.\nonumber \\ && \Big.F\Big(-\ell-1,-\ell+\omega L,-2\ell;\dfrac{2r}{r+iL}\Big)\Big]\;,\label{farsol}
\end{eqnarray}
where $C_1$, $C_2$ are two integration constants, and they will be constrained in the following, in order to satisfy the boundary conditions.

The first boundary condition in Eq.~\eqref{bc1}, in the far region, becomes
\begin{equation}
\dfrac{\alpha^{-}}{\beta^{-}}=\dfrac{i}{\omega L^2}\;.\nonumber
\end{equation}
In order to impose this boundary condition, we first expand Eq.~\eqref{farsol} at large $r$, in the form of Eq.~\eqref{asysol}; then one obtains the first relation between $C_1$ and $C_2$
\begin{equation}
\dfrac{C_2}{C_1}=-2^{2\ell+1}\dfrac{\ell}{\ell+1}\dfrac{F(\ell+1,\ell+1+\omega L,2\ell+2;2)}{F(-\ell,-\ell+\omega L,-2\ell;2)}\;.\label{c1c2rel1}
\end{equation}
The second boundary condition in Eq.~\eqref{bc2}, in the far region, turns to
\begin{equation}
\dfrac{\alpha^{-}}{\beta^{-}}=\dfrac{i\omega}{-\ell(\ell+1)+\omega^2L^2}\;.\nonumber
\end{equation}
To impose the second boundary condition above, again expanding Eq.~\eqref{farsol} at large $r$, to extract $\alpha^{-}$ and $\beta^{-}$, then one gets the second relation between $C_1$ and $C_2$
\begin{equation}
\dfrac{C_2}{C_1}=2^{2\ell+1}\left(\dfrac{\ell}{\ell+1}\right)^2\dfrac{\ell+1+\omega L}{\ell-\omega L}\dfrac{\mathcal{A}_1}{\mathcal{A}_2}\;,
\end{equation}
where
\begin{eqnarray}
\mathcal{A}_1=&&(\ell+1)F(\ell,\ell+1+\omega L,2\ell+2;2)+\omega L F(\ell+1,\nonumber\\&&\ell+2+\omega L,2\ell+3;2)\;,\nonumber\\
\mathcal{A}_2=&&\ell F(-\ell-1,-\ell+\omega L,-2\ell;2)-\omega L F(-\ell,-\ell+1\nonumber\\&&+\omega L,1-2\ell;2)\;.
\end{eqnarray}
In order to match this solution to the near region solution, we expand Eq.~\eqref{farsol} for small $r$, to obtain
\begin{equation}
R_{-1}\;\sim\;\dfrac{R^{\rm far}_{-1,1/r}}{r^\ell}+R^{\rm far}_{-1,r}r^{\ell+1}\;,\label{farsolnear}
\end{equation}
with
\begin{eqnarray}
&&R^{\rm far}_{-1,1/r}  \equiv   -i L C_2 \ ,\nonumber \\
&&R^{\rm far}_{-1,r} \equiv (-1)^\ell2^{2\ell+1}L^{-2\ell} C_1\;.\nonumber
\end{eqnarray}

\subsection{Overlap region}
To match the near region solution Eq.~\eqref{nearsolfar} and the far region solution Eq.~\eqref{farsolnear} in the intermediate region, we impose the matching condition $R^{\rm near}_{-1,r}R^{\rm far}_{-1,1/r}=R^{\rm far}_{-1,r}R^{\rm near}_{-1,1/r}$. Then we can get
\begin{eqnarray}
&&\dfrac{\Gamma(-2\ell-1)}{\Gamma(-\ell)}\dfrac{\Gamma(\ell+1)}{\Gamma(2\ell+1)}\dfrac{\Gamma(\ell+2-2i\bar{\omega})}{\Gamma(1-\ell-2i\bar{\omega})}\left(\dfrac{r_+-r_-}{L}\right)^{2\ell+1}
\nonumber\\
&&=\;i\;(-1)^\ell \dfrac{\ell}{\ell+1}\dfrac{F(\ell+1,\ell+1+\omega L,2\ell+2;2)}{F(-\ell,-\ell+\omega L,-2\ell;2)}\;,\label{match1}
\end{eqnarray}
with the first boundary condition given by Eq.~\eqref{bc1}, and
\begin{eqnarray}
&&\dfrac{\Gamma(-2\ell-1)}{\Gamma(-\ell)}\dfrac{\Gamma(\ell+1)}{\Gamma(2\ell+1)}\dfrac{\Gamma(\ell+2-2i\bar{\omega})}{\Gamma(1-\ell-2i\bar{\omega})}\left(\dfrac{r_+-r_-}{L}\right)^{2\ell+1}
\nonumber\\
&&=\;i\;(-1)^{\ell+1}\left(\dfrac{\ell}{\ell+1}\right)^2\dfrac{\ell+1+\omega L}{\ell-\omega L}\;\dfrac{\mathcal{A}_1}{\mathcal{A}_2}\;,\label{match2}
\end{eqnarray}
with the second boundary condition given by Eq.~\eqref{bc2}.
Both Eqs.~\eqref{match1} and~\eqref{match2} can be solved perturbatively to look for the imaginary part of quasinormal frequencies, in the small BH $(r_+\ll L)$ and slow rotation $(a\ll r_+)$ approximations.
In order to do so, we first look for normal modes. For a small BH, the left term in Eqs.~\eqref{match1} and~\eqref{match2} vanish at the leading order, then we have to require the right term in both equations to vanish as well. These conditions give the normal modes for pure AdS
\begin{eqnarray}
&&F(\ell+1,\ell+1+\omega L,2\ell+2;2)=0\;\nonumber\\
&&\Rightarrow\;\;\omega_{1,N}L=2N+\ell+2\;,\label{normalmode1}\\
&&\mathcal{A}_1=0\;\nonumber\\
&&\Rightarrow\;\;\omega_{2,N}L=2N+\ell+1\;,\label{normalmode2}
\end{eqnarray}
where $N=0,1,2,\cdot\cdot\cdot$, and $\ell=1,2,3,\cdot\cdot\cdot$. The two sets of modes are, in this case, isospectral up to one mode~\cite{Wang:2015goa}.

When the BH effects are taken into account, a correction to the frequency will be introduced
\begin{equation}
\omega_j L=\omega_{j,N} L+i\delta_j\;,\label{normalmode}
\end{equation}
where $j=1,2$ for the two different boundary \mbox{conditions}, and $\delta$ is used to describe the damping (growth) of the quasinormal modes, and we replace $\omega L$ appearing in the second line of Eqs.~\eqref{match1} and~\eqref{match2} by $\omega_1 L$ and $\omega_2 L$ in \mbox{Eq.~\eqref{normalmode}}, respectively. Then, from each of these two \mbox{equations}, we can obtain $\delta_j$ perturbatively, at leading order in $a$.

It turns out that  the general expression for $\delta_j$ is quite messy. As such, we only show here a few explicit examples. For $\ell=1$ and $N=0$ case, from Eq.~\eqref{match1}, we get
\begin{eqnarray}
\delta_1&=&-\dfrac{16}{\pi}\dfrac{r_+^4}{L^4}+m\dfrac{16}{3\pi}\dfrac{ar_+^2}{L^3}+\mathcal{O}\left(\frac{a}{L},\frac{r_+^4}{L^4}\right)\nonumber\\
&=&-\dfrac{16}{3\pi}\dfrac{r_+^2}{L^2}\left(3\dfrac{r_+^2}{L^2}-m\dfrac{a}{L}\right)+\cdot\cdot\cdot\nonumber\\
&\simeq&-\dfrac{16}{3\pi}\dfrac{r_+^4}{L^3}\left(\omega_{1,0}-m\Omega_H\right)+\cdot\cdot\cdot\;,\label{analy1}
\end{eqnarray}
where the angular velocity has been approximated by $\Omega_H\sim a/r_+^2$. It is manifest, from \eqref{analy1}, that $\delta_1<0$ when $\omega_{1,0}>m\Omega_H$, while $\delta_1>0$ when $\omega_{1,0}<m\Omega_H$. Thus we find growing modes within the superradiant regime, as expected.
Keeping the same parameters as in the previous paragraph, $i.e.$, $\ell=1$ and $N=0$, from Eq.~\eqref{match2}, we obtain
\begin{eqnarray}
\delta_2&=&-\dfrac{8}{3\pi}\dfrac{r_+^4}{L^4}+m\dfrac{4}{3\pi}\dfrac{ar_+^2}{L^3}+\mathcal{O}\left(\frac{a}{L},\frac{r_+^4}{L^4}\right)\nonumber\\
&=&-\dfrac{4}{3\pi}\dfrac{r_+^2}{L^2}\left(2\dfrac{r_+^2}{L^2}-m\dfrac{a}{L}\right)+\cdot\cdot\cdot\nonumber\\
&\simeq&-\dfrac{4}{3\pi}\dfrac{r_+^4}{L^3}\left(\omega_{2,0}-m\Omega_H\right)+\cdot\cdot\cdot\;,\label{analy2}
\end{eqnarray}
which also shows clearly that $\delta_2<0$ when $\omega_{2,0}>m\Omega_H$, but $\delta_2>0$ when $\omega_{2,0}<m\Omega_H$, signaling again superradiant instabilties.

Furthermore, for both cases, $\delta_1=0$ and $\delta_2=0$ give $\omega_{1,0}=m\Omega_H$ and $\omega_{2,0}=m\Omega_H$, which are the conditions to form clouds, with the two different boundary conditions.

\section{Numerics}
\label{secnumerical}
When the BH parameters lie beyond the small and \mbox{slow} rotation approximations provided in the last section, the analytical method fails and we have to solve the problem numerically. Since the radial equation~\eqref{radialeq} and the angular equation~\eqref{angulareq} are coupled through their eigenvalues, we have to solve both equations simultaneously. In this section, we will first demonstrate the numerical method applied in this paper, and then present some numerical results.

\subsection{Method}
\label{NM}
The radial equation~\eqref{radialeq}, will be solved by the \mbox{direct} integration method, adapted from our previous \mbox{works}~\cite{Herdeiro:2011uu,Wang:2012tk,Sampaio:2014swa}. For a self-contained presentation, we briefly outline the procedure here.
We first use Frobenius' method to expand $R_{-1}$ close to the event horizon
\begin{equation}
R_{-1}=(r-r_+)^\rho \sum_{j=0}^\infty c_j\;(r-r_+)^j\;,\nonumber
\end{equation}
to initialize Eq.~\eqref{radialeq}. The series expansion coefficients $c_j$ can be derived directly after inserting these expansions into Eq.~\eqref{radialeq}.
The parameter $\rho$ is chosen as
\begin{equation}
\rho=1-\dfrac{i(\omega-m\Omega_H)}{4\pi T_H}\;,\nonumber
\end{equation}
so that the ingoing boundary condition is satisfied. The angular velocity $\Omega_H$ and the Hawking temperature $T_H$ are given in Eq.~\eqref{Horvel} and Eq.~\eqref{Temp}.
At infinity, the asymptotic behavior of $R_{-1}$ is given by Eq.~\eqref{asysol}. The expansion coefficients, $\alpha^{-}$ and $\beta^{-}$, can be extracted from $R_{-1}$ and its first derivative. For that purpose, we define two new fields $\left\{\chi,\psi\right\}$, which asymptote respectively to $\left\{\alpha^{-},\beta^{-}\right\}$ at infinity. Such a transformation can be written in the matrix form
\begin{equation}
\mathbf{V}= \left(\begin{array}{cc} r & 1 \vspace{2mm}\\ 1 & 0 \end{array}\right) \mathbf{\Psi} \equiv \mathbf{T} \mathbf{\Psi} \;,\nonumber
\end{equation}
by defining the vector $\mathbf{\Psi}^T=(\chi,\psi)$ for the new fields, and another vector $\mathbf{V}^T=(R_{-1},\frac{d}{dr}R_{-1})$ for the original field and its derivative.
To obtain a first order system of ODEs for the new fields, we define another matrix $\mathbf{X}$, through
\begin{equation}
\dfrac{d\mathbf{V}}{dr}=\mathbf{X}\mathbf{V} \; ,
\end{equation}
which can be read off from the original radial equation~\eqref{radialeq} directly.
Then the radial equation~\eqref{radialeq} becomes
\begin{equation}
\dfrac{d\mathbf{\Psi}}{dr}=\mathbf{T}^{-1}\left(\mathbf{X}\mathbf{T}-\dfrac{d\mathbf{T}}{dr}\right) \mathbf{\Psi} \; .\label{radialmatrix}
\end{equation}
This is the final equation we are going to solve.

The angular equation~\eqref{angulareq}, will be solved by a spectral method, to look for the separation constant $\lambda$. By observing Eq.~\eqref{angulareq} and considering the constraint on rotation, $a\ll L$, one finds two regular singularities, at $u=\pm 1$. To impose regular boundary conditions at these regular singularities, we require
\begin{equation}
S\;\sim\;
\begin{cases}
(1-u)^{\frac{|m+s|}{2}}\;\;\;\;\;\text{when}\;\;\;u\rightarrow 1\;,\\(1+u)^{\frac{|m-s|}{2}}\;\;\;\;\;\text{when}\;\;\;u\rightarrow -1\;,
\end{cases}
\end{equation}
where, as announced before, $s=-1$. These asymptotic solutions can be factored out by defining a new function $\hat{S}$
\begin{equation}
S=(1-u)^{\frac{|m+s|}{2}}(1+u)^{\frac{|m-s|}{2}}\hat{S}\; .\label{trans}
\end{equation}
Then the angular equation~\eqref{angulareq} becomes
\begin{equation}
Y(u)\hat{S}=\lambda \hat{S}\;,\label{eigenfunc}
\end{equation}
where the operator $Y(u)$ can be obtained straightforwardly after inserting the transformation~\eqref{trans} into the angular equation~\eqref{angulareq}.
We choose Chebyshev grids as the collocation points to discretize the operator $Y(u)$, which turns out to be a matrix. Then Eq~\eqref{eigenfunc} becomes a linear algebraic equation, and $\lambda$ is obtained by looking for the eigenvalues of the matrix $Y(u)$.

\subsection{Results}
With the numerical strategies described above, and the boundary conditions given in Eqs.~\eqref{bc1} and~\eqref{bc2}, the eigenvalues $\{\omega,\lambda\}$ of the coupled system in Eqs.~\eqref{radialmatrix} and~\eqref{eigenfunc} can be solved iteratively, through assuming an initial guess for $\omega$ or $\lambda$, until solutions $\{\omega,\lambda\}$ become stable. The initial values for $\omega$ or $\lambda$ can be chosen from the results in Schwarzschild-AdS BHs~\cite{Wang:2015goa} or $\ell(\ell+1)$.

Note that all the physical quantities in the numerical calculations are normalized by the AdS radius $L$ and we set $L=1$. Also note that we use $\omega_1$ $(\omega_2)$ to represent the quasinormal frequency and $\lambda_1$ ($\lambda_2$) to stand for the separation constant, corresponding to the first (second) boundary conditions.

\subsubsection{Quasinormal modes and superradiant instabilities}
A few selected eigenvalues of $\omega$ and $\lambda$ are tabulated in Tables~\ref{Table1}$-$\ref{Table3}, with the two boundary conditions, for various BH sizes. Since the superradiant instability is one of our main interests, and it is a generic feature that lower order modes exhibit a stronger instability, we focus on the lowest fundamental modes, characterized by $N=0,\;\ell=1$ and $m=0,\;\pm1$.

In Table~\ref{Table1} we consider a small BH with size $r_+=0.1$. The first observation from this table is that superradiant instabilities exist for both boundary conditions, with positive $m$; this is because only positive $m$ modes can meet the superradiance condition, assuming positive frequencies.

\begin{table*}
\centering
\caption{\label{Table1} Quasinormal frequencies and separation constants of the Maxwell field with the two different boundary conditions, for $\ell=1$ fundamental modes, on a Kerr-AdS BH with size $r_+=0.1$.}
\begin{ruledtabular}
\begin{tabular}{ l l l l l l }
$(\ell,m)$ & $a$ & $\omega_1$ & $\lambda_1$ & $\omega_2$ & $\lambda_2$ \\
\hline\\
(1,\;0) & 0 & 2.8519 - 1.7050$\times 10^{-3}$ i & 2 & 1.9533 - 1.8240$\times 10^{-4}$ i & 2\\
  & 0.01 & 2.8505 - 1.7295$\times 10^{-3}$ i & 2.0005 - 5.9147$\times 10^{-7}$ i & 1.9529 - 1.8329$\times 10^{-4}$ i & 2.0003 - 4.2946$\times 10^{-8}$ i\\
  & 0.05 & 2.8151 - 2.5818$\times 10^{-3}$ i & 2.0133 - 2.1677$\times 10^{-5}$ i & 1.9452 - 2.0829$\times 10^{-4}$ i & 2.0071 - 1.2092$\times 10^{-6}$ i\\
  & 0.1 & 2.6740 - 2.2847$\times 10^{-2}$ i & 2.0480 - 7.1650$\times 10^{-4}$ i & 1.9160 - 1.0036$\times 10^{-3}$ i & 2.0276 - 2.2602$\times 10^{-5}$ i\\
  \hline\\
(1,\;1) & 0.01 & 2.8436 - 9.5962$\times 10^{-4}$ i & 1.9149 + 2.8541$\times 10^{-5}$ i & 1.9436 - 7.9809$\times 10^{-5}$ i & 1.9417 + 2.3800$\times 10^{-6}$ i\\
  & 0.05 & 2.7837 + 5.5800$\times 10^{-4}$ i & 1.5879 - 8.0057$\times 10^{-5}$ i & 1.8989 + 1.9474$\times 10^{-4}$ i & 1.7156 - 2.8302$\times 10^{-5}$ i\\
  & 0.1 & 2.6493 + 2.0481$\times 10^{-3}$ i & 1.2278 - 5.6148$\times 10^{-4}$ i & 1.8292 + 7.8282$\times 10^{-4}$ i & 1.4552 - 2.1958$\times 10^{-4}$ i\\
  \hline\\
  (1,\;-1) & 0.01 & 2.8572 - 2.7984$\times 10^{-3}$ i & 2.0859 - 8.4666$\times 10^{-5}$ i & 1.9622 - 3.2541$\times 10^{-4}$ i & 2.0589 - 9.8188$\times 10^{-6}$ i\\
  & 0.05 & 2.8422 - 1.7571$\times 10^{-2}$ i & 2.4296 - 2.7419$\times 10^{-3}$ i & 1.9900 - 2.1380$\times 10^{-3}$ i & 2.2975 - 3.2933$\times 10^{-4}$ i\\
  & 0.1 & 2.7398 - 1.2611$\times 10^{-1}$ i & 2.8269 - 4.0538$\times 10^{-2}$ i & 1.9987 - 1.9186$\times 10^{-2}$ i & 2.5914 - 6.0286$\times 10^{-3}$ i
\end{tabular}
\end{ruledtabular}
\end{table*}

\begin{figure*}
\begin{center}
\begin{tabular}{c}
\hspace{-4mm}\includegraphics[clip=true,width=0.38\textwidth]{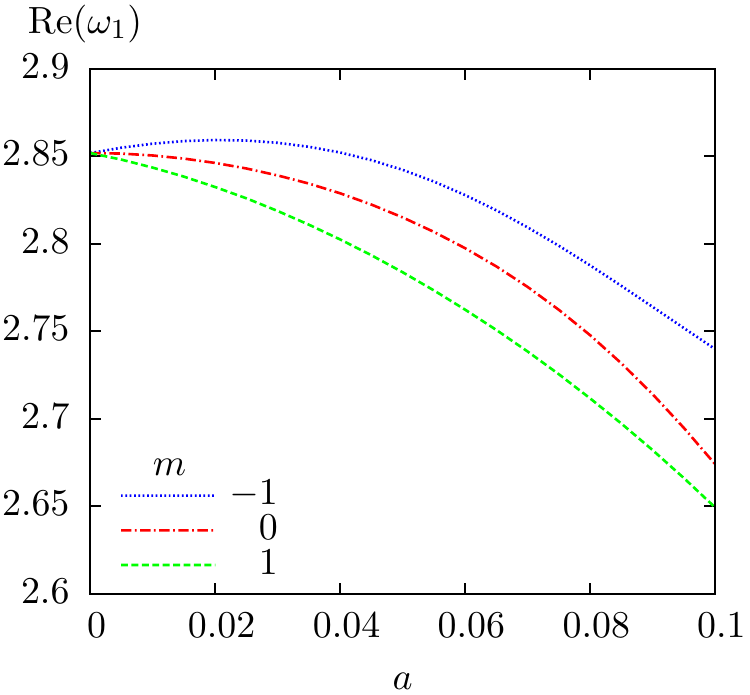}\hspace{6mm}\includegraphics[clip=true,width=0.38\textwidth]{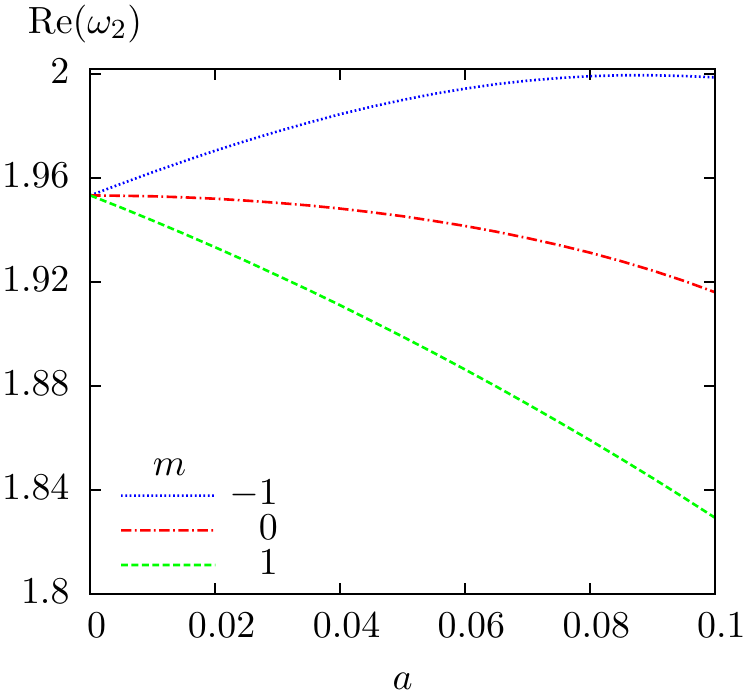}
\end{tabular}
\end{center}
\caption{\label{rp01refre} Variation of Re($\omega$) with varying rotation parameter, for fixed $r_+=0.1$ and $\ell=1$ but for different values $m$. The left panel is for the first boundary condition while the right panel is for the second boundary condition.}
\end{figure*}

\begin{figure*}
\begin{center}
\begin{tabular}{c}
\hspace{-4mm}\includegraphics[clip=true,width=0.39\textwidth]{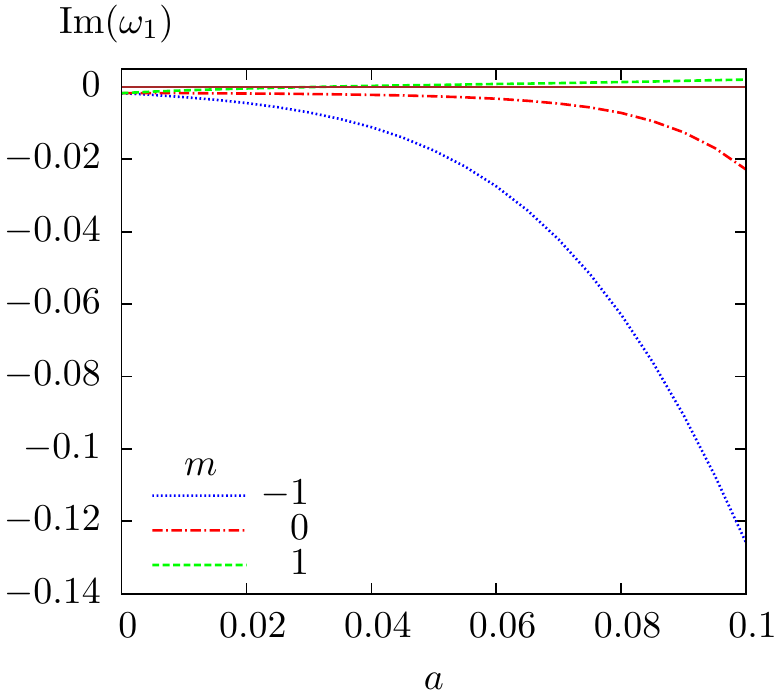}\hspace{6mm}\includegraphics[clip=true,width=0.39\textwidth]{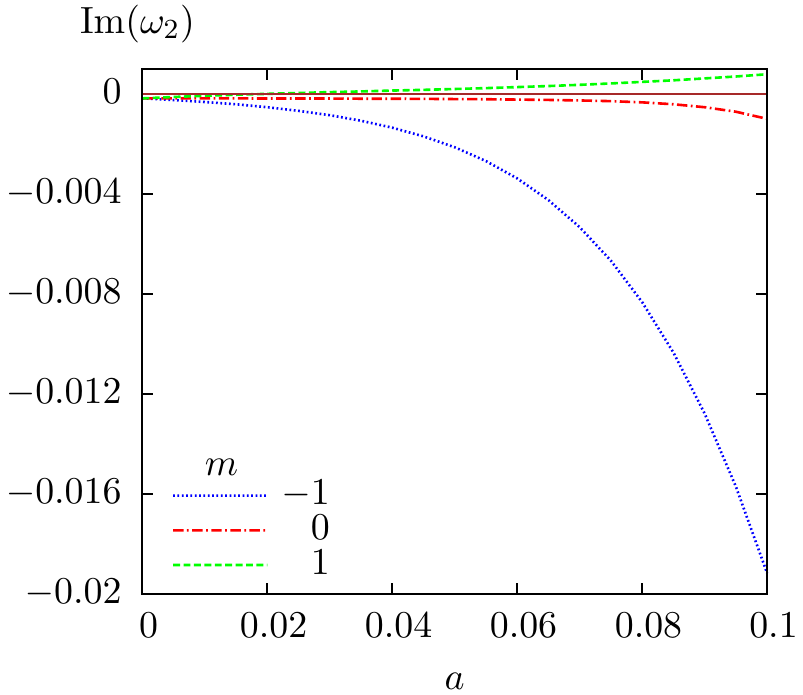}
\end{tabular}
\end{center}
\caption{\label{rp01imfre} Variation of Im($\omega$) with varying rotation parameter, for fixed $r_+=0.1$ and $\ell=1$ but for different values $m$. The left panel is for the first boundary condition while the right panel is for the second boundary condition. The brown solid thin line corresponds to Im($\omega$)=0, to exhibit more clearly  superradiant instabilities.}
\end{figure*}

\begin{figure*}
\begin{center}
\begin{tabular}{c}
\hspace{-4mm}\includegraphics[clip=true,width=0.38\textwidth]{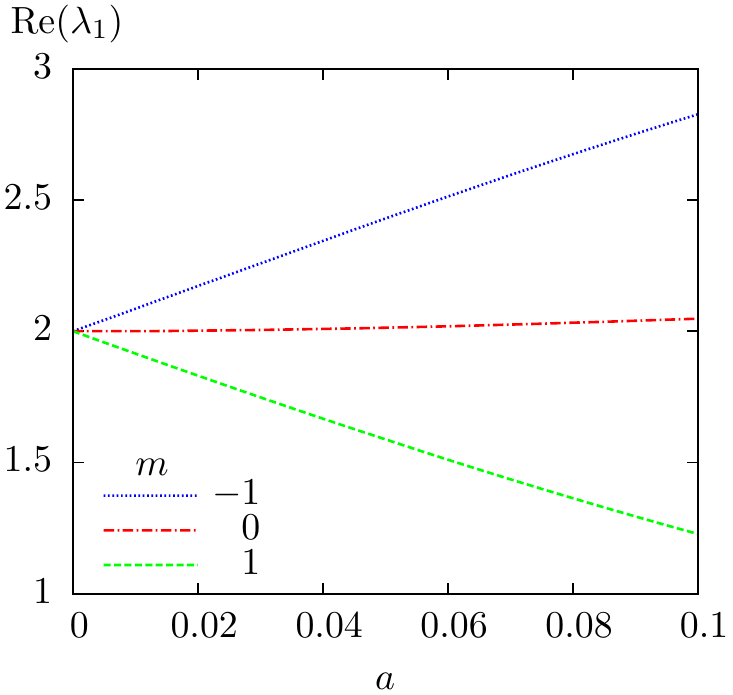}\hspace{6mm}\includegraphics[clip=true,width=0.38\textwidth]{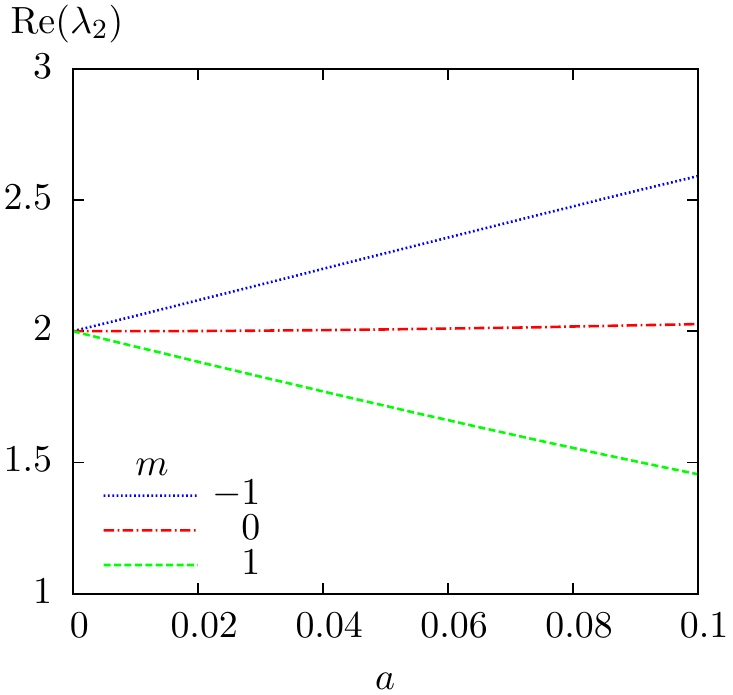}
\end{tabular}
\end{center}
\caption{\label{rp01relam} Variation of Re($\lambda$) with varying rotation parameter, for fixed $r_+=0.1$ and $\ell=1$ but for different values $m$. The left panel is for the first boundary condition while the right panel is for the second boundary condition.}
\end{figure*}

\begin{figure*}
\begin{center}
\begin{tabular}{c}
\hspace{-4mm}\includegraphics[clip=true,width=0.39\textwidth]{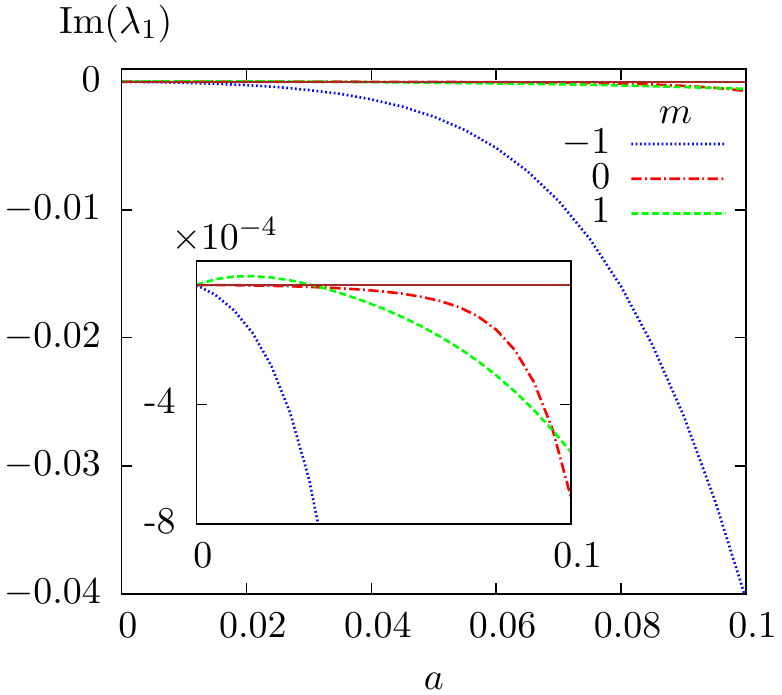}\hspace{6mm}\includegraphics[clip=true,width=0.39\textwidth]{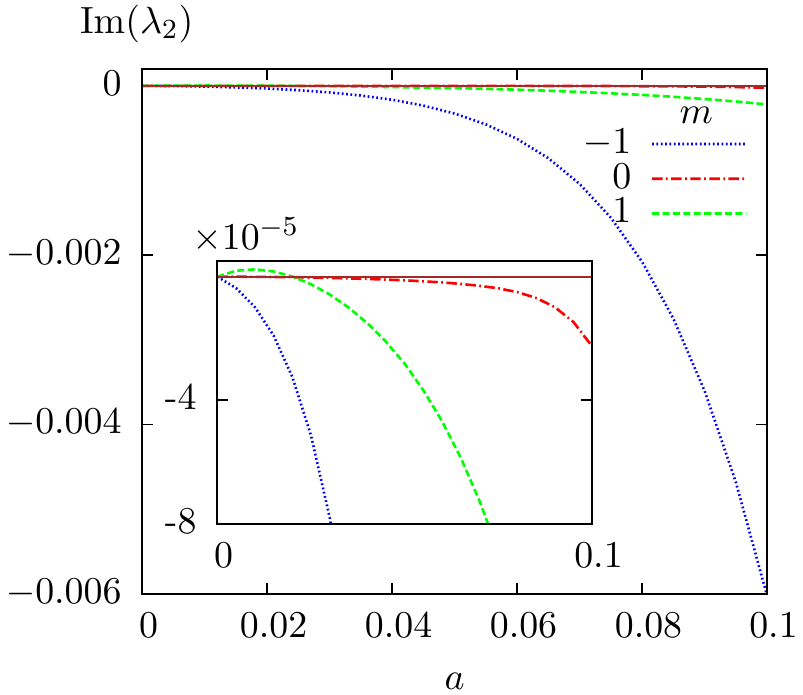}
\end{tabular}
\end{center}
\caption{\label{rp01imlam} Variation of Im($\lambda$) with varying rotation parameter, for fixed $r_+=0.1$ and $\ell=1$ but for different values $m$. The left panel is for the first boundary condition while the right panel is for the second boundary condition. Again, the brown solid thin line corresponds to Im($\lambda$)=0, to exhibit that the sign of Im($\lambda$) changes when the superradiant instability occurs (seen in the insets).}
\end{figure*}

The effect of varying the rotation parameter on both eigenvalues, for different values of $m$ with fixed $\ell=1$, are shown in Figs.~\ref{rp01refre}$-$\ref{rp01imlam}. In Fig.~\ref{rp01refre}, the real part of the frequency is shown. An immediate first impression from Fig.~\ref{rp01refre}, is that it seems the rotation impacts differently on the $m=-1$ modes, for the two boundary conditions. Checking carefully the numerical data for Re($\omega_2$), however, we find that its value decrease sightly when $a$ is approaching $0.1$. Thus,  for  $m=-1$ and for both boundary conditions, Re($\omega$) starts by increasing with increasing rotation but then decreases. For the other two values of $m$, Re($\omega$) always decreases with increasing rotation.

In Fig.~\ref{rp01imfre}, the imaginary part of the frequency is shown, for both boundary conditions. Im($\omega$) increases with increasing rotation when $m=1$, eventually becoming positive, signaling the presence of superradiant unstable modes.

 In Fig.~\ref{rp01relam} and Fig.~\ref{rp01imlam}, we show the real and imaginary part of the separation constant, respectively, for both boundary conditions. The real part increases with increasing rotation both when $m=-1$ and $m=0$, albeit only slightly in the latter case, and decreases with the rotation when $m=1$. As for the imaginary part, for $m=1$, it increases with the rotation initially in a small range; but it starts decreasing afterwards. For the other two values of $m$, the imaginary part of the separation constants always decrease with the rotation. We also note that Im($\lambda$), for $m=0$, with the first boundary condition decays faster than its counterpart with the second boundary condition. From these first four figures, we conclude that the effect of varying the rotation on the eigenvalues is similar for the two boundary conditions. In the following, therefore, we only show the rotation effect on the eigenvalues with the first boundary condition, when considering other BH sizes.


\begin{figure*}
\begin{center}
\begin{tabular}{c}
\hspace{-4mm}\includegraphics[clip=true,width=0.38\textwidth]{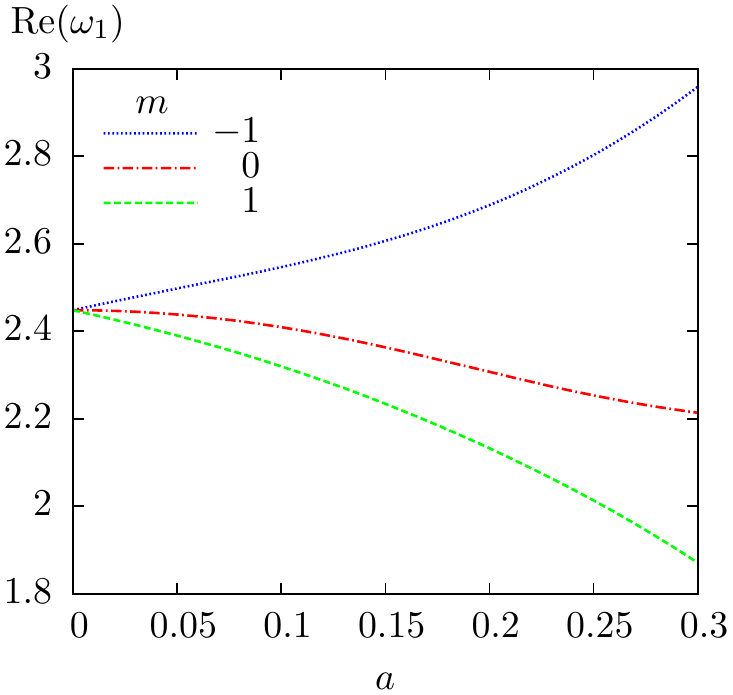}\hspace{6mm}\includegraphics[clip=true,width=0.39\textwidth]{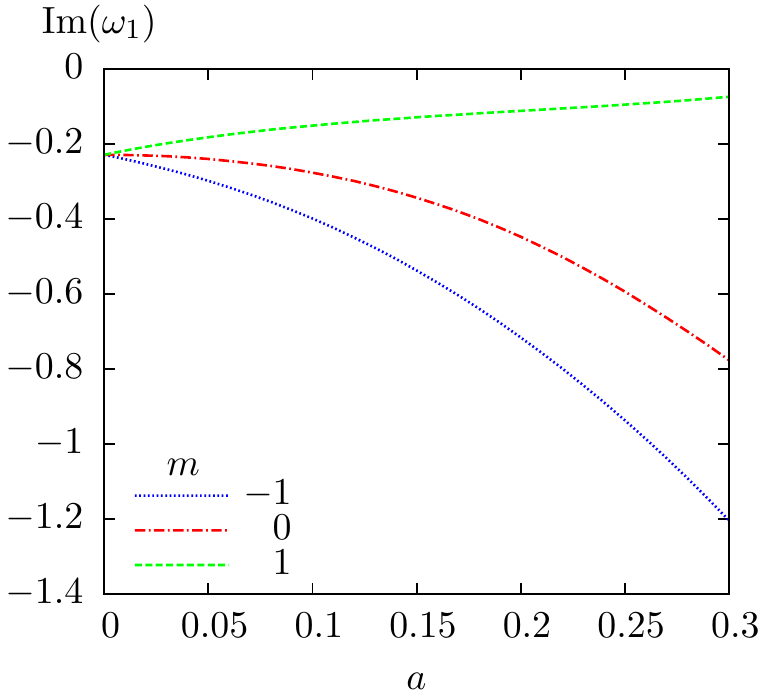}
\end{tabular}
\end{center}
\caption{\label{rp03fre} Variation of $\omega$ with varying rotation parameter, for different values of $m$. The BH size is fixed as $r_+=0.3$ and the first boundary condition has been imposed. The left panel is for Re($\omega$) while the right panel is for Im($\omega$).}
\end{figure*}

\begin{figure*}
\begin{center}
\begin{tabular}{c}
\hspace{-4mm}\includegraphics[clip=true,width=0.38\textwidth]{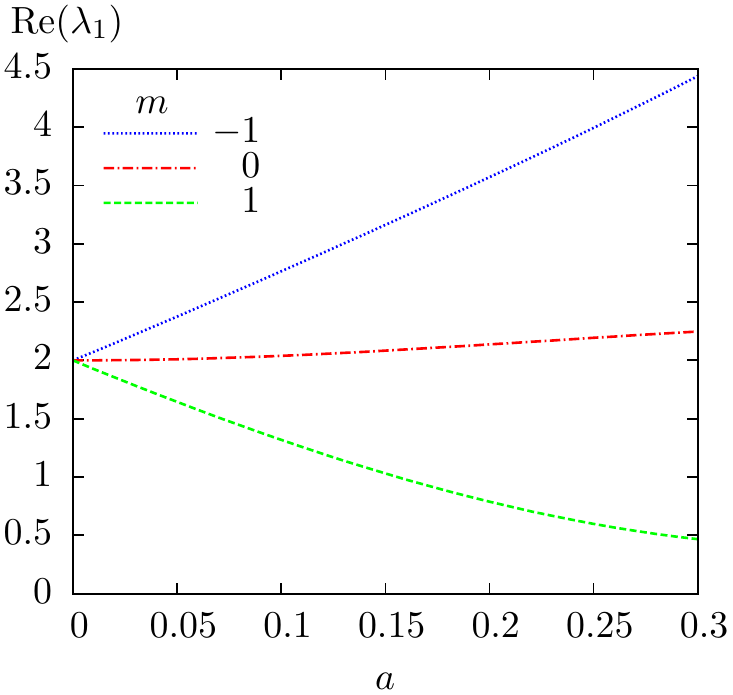}\hspace{6mm}\includegraphics[clip=true,width=0.39\textwidth]{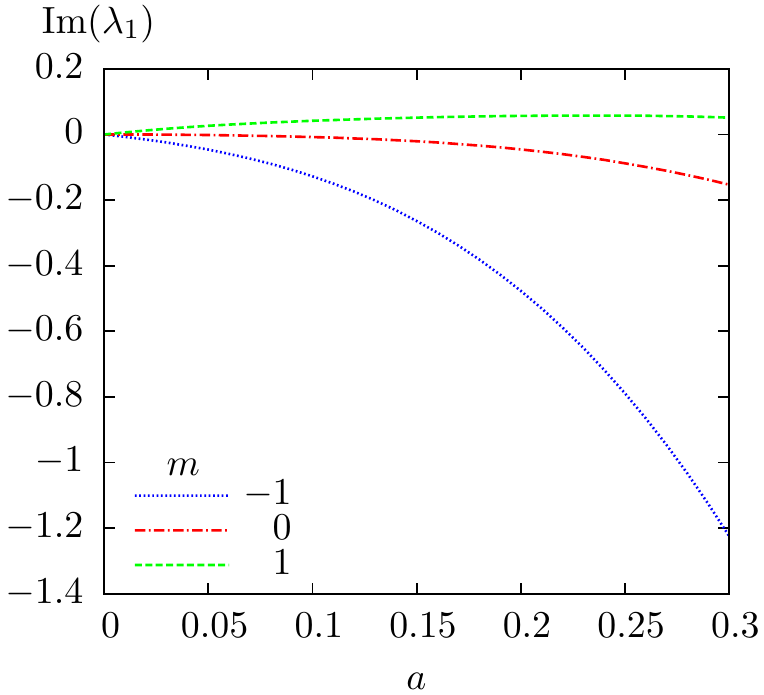}
\end{tabular}
\end{center}
\caption{\label{rp03lam} Variation of $\lambda$ with varying rotation parameter, for different values of $m$. The BH size is fixed as $r_+=0.3$ and the first boundary condition has been imposed. The left panel is for Re($\lambda$) while the right panel is for Im($\lambda$).}
\end{figure*}

We continue our study by varying the BH size. In Table~\ref{Table2} we list a few selected eigenvalues for $r_+=0.3$. The interesting feature that now emerges is that superradiant instabilities only occur for the second boundary condition. This implies that the second boundary condition may produce unstable modes in a larger parameter space. This feature will be shown more clearly in the parameter space for the vector clouds. The effect of varying the rotation on the eigenvalues is shown in Figs.~\ref{rp03fre}$-$\ref{rp03lam}, with the first boundary condition. In Fig.~\ref{rp03fre}, it displays that, Re($\omega$) increases with increasing rotation for $m=-1$ mode but decreases for both $m=0$ and $m=1$ modes, while Im($\omega$) increases with increasing rotation for $m=1$ mode but decreases for both $m=0$ and $m=-1$ modes. Behaviors of the separation constant, shown in Fig.~\ref{rp03lam}, are similar to the counterparts in $r_+=0.1$ case\footnote{Notice that for $m=1$ mode, Im($\lambda$) starts decreasing with rotation around $a=0.25$.}.

\begin{table*}
\centering
\caption{\label{Table2} Quasinormal frequencies and separation constants of the Maxwell field with the two different boundary conditions, for $\ell=1$ fundamental modes, on a Kerr-AdS BH with size $r_+=0.3$.}
\begin{ruledtabular}
\begin{tabular}{ l l l l l l }
$(\ell,m)$ & $a$ & $\omega_1$ & $\lambda_1$ & $\omega_2$ & $\lambda_2$ \\
\hline\\
(1,\;0) & 0 & 2.4481 - 0.2291 i & 2 & 1.8152 - 3.8034$\times 10^{-2}$ i & 2\\
  & 0.1 & 2.4093 - 0.2768 i & 2.0397 - 7.8286$\times 10^{-3}$ i & 1.8092 - 4.7641$\times 10^{-2}$ i & 2.0252 - 1.0134$\times 10^{-3}$ i\\
  & 0.2 & 2.3071 - 0.4480 i & 2.1373 - 4.5479$\times 10^{-2}$ i & 1.7921 - 8.8195$\times 10^{-2}$ i & 2.0949 - 6.9871$\times 10^{-3}$ i\\
  & 0.3 & 2.2136 - 0.7769 i & 2.2476 - 1.5300$\times 10^{-1}$ i & 1.7875 - 1.8395$\times 10^{-1}$ i & 2.1957 - 2.9422$\times 10^{-2}$ i\\
  \hline\\
(1,\;1) & 0.1 & 2.3197 - 0.1512 i & 1.3185 + 4.1838$\times 10^{-2}$ i & 1.7265 - 1.5809$\times 10^{-2}$ i & 1.4841 + 4.4473$\times 10^{-3}$ i\\
  & 0.2 & 2.1325 - 0.1120 i & 0.7889 + 5.6598$\times 10^{-2}$ i & 1.6328 - 2.8939$\times 10^{-3}$ i & 1.0447 + 1.4994$\times 10^{-3}$ i\\
  & 0.3 & 1.8707 - 7.4377$\times 10^{-2}$ i & 0.4681 + 5.1126$\times 10^{-2}$ i & 1.5304 + 5.2612$\times 10^{-3}$ i & 0.7048 - 3.6993$\times 10^{-3}$ i\\
  \hline\\
  (1,\;-1) & 0.1 & 2.5468 - 0.3994 i & 2.7643 - 0.1276 i & 1.9063 - 8.8702$\times 10^{-2}$ i & 2.5624 - 2.7793$\times 10^{-2}$ i\\
  & 0.2 & 2.6884 - 0.7162 i & 3.5705 - 0.4764 i & 2.0270 - 1.9387$\times 10^{-1}$ i & 3.1487 - 1.2392$\times 10^{-1}$ i\\
  & 0.3 & 2.9588 - 1.2044 i & 4.4377 - 1.2239 i & 2.2192 - 3.6493$\times 10^{-1}$ i & 3.7662 - 3.4830$\times 10^{-1}$ i
\end{tabular}
\end{ruledtabular}
\end{table*}

\begin{figure*}
\begin{center}
\begin{tabular}{c}
\hspace{-4mm}\includegraphics[clip=true,width=0.38\textwidth]{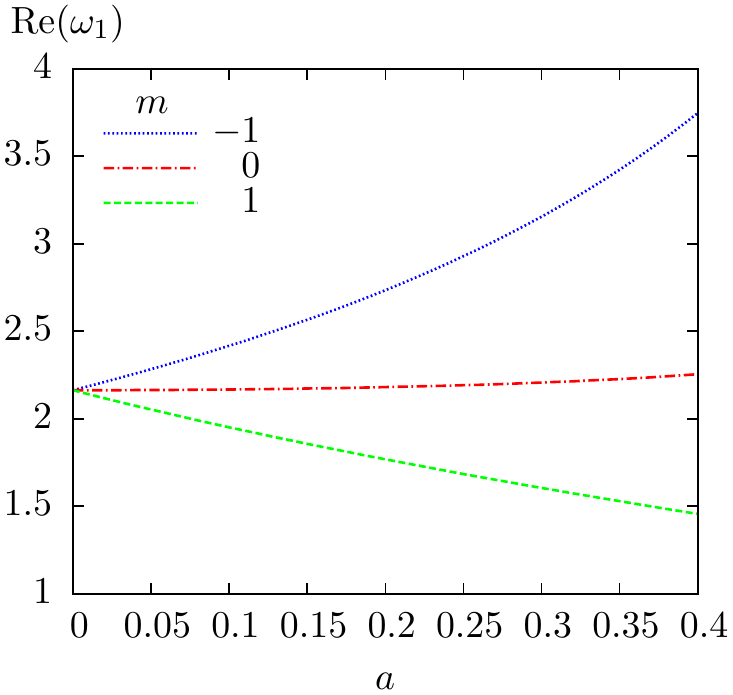}\hspace{6mm}\includegraphics[clip=true,width=0.39\textwidth]{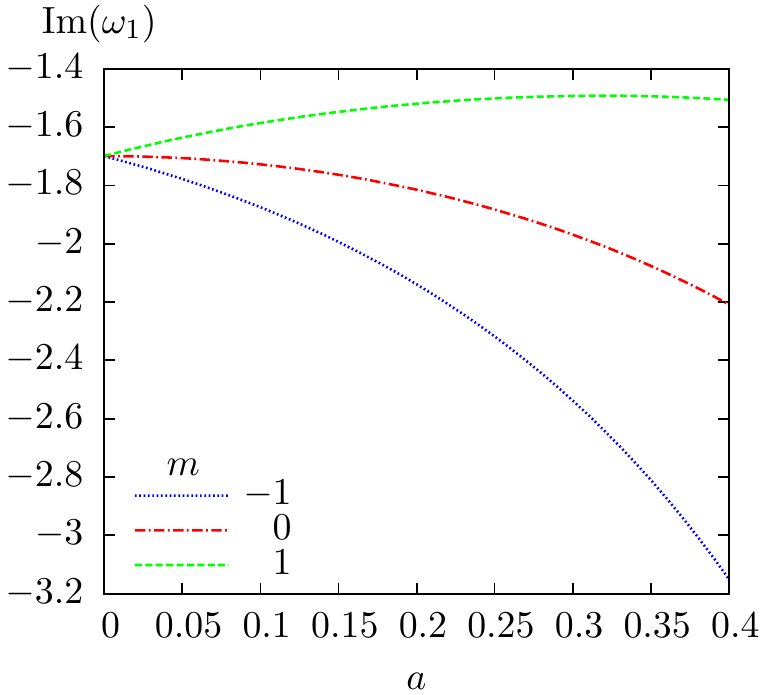}
\end{tabular}
\end{center}
\caption{\label{rp1fre} Variation of $\omega$ with varying rotation parameter, for different values of $m$. The BH size is fixed as $r_+=1$ and the first boundary condition has been imposed. The left panel is for Re($\omega$) while the right panel is for Im($\omega$).}
\end{figure*}

\begin{figure*}
\begin{center}
\begin{tabular}{c}
\hspace{-4mm}\includegraphics[clip=true,width=0.38\textwidth]{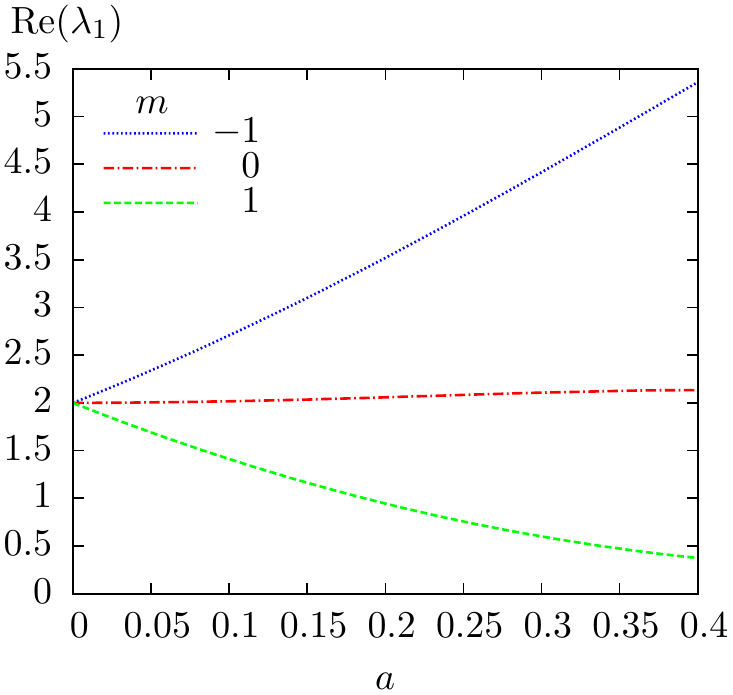}\hspace{6mm}\includegraphics[clip=true,width=0.38\textwidth]{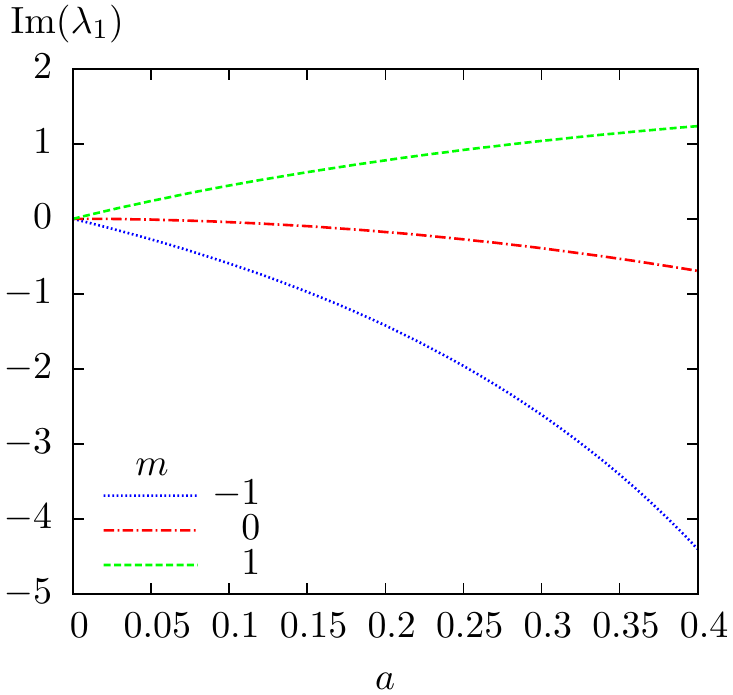}
\end{tabular}
\end{center}
\caption{\label{rp1lam} Variation of $\lambda$ with varying rotation parameter, for different values of $m$. The BH size is fixed as $r_+=1$ and the first boundary condition has been imposed. The left panel is for Re($\lambda$) while the right panel is for Im($\lambda$).}
\end{figure*}

The results for $r_+=1$ are presented in Table~\ref{Table3} and Figs.~\ref{rp1fre}$-$\ref{rp1lam}. From Table~\ref{Table3} one observes there is no superradiant instability for any of the boundary conditions.

\begin{table*}
\centering
\caption{\label{Table3} Quasinormal frequencies and separation constants of the Maxwell field with the two different boundary conditions, for $\ell=1$ fundamental modes, on a Kerr-AdS BH with size $r_+=1$.}
\begin{ruledtabular}
\begin{tabular}{ l l l l l l }
$(\ell,m)$ & $a$ & $\omega_1$ & $\lambda_1$ & $\omega_2$ & $\lambda_2$ \\
\hline\\
(1,\;0) & 0 & 2.1630 - 1.6991 i & 2 & 1.5536 - 0.5418 i & 2\\
  & 0.1 & 2.1672 - 1.7274 i & 2.0162 - 4.4059$\times 10^{-2}$ i & 1.5627 - 0.5510 i & 2.0186 - 1.0131$\times 10^{-2}$ i\\
  & 0.2 & 2.1805 - 1.8146 i & 2.0580 - 0.1757 i & 1.5909 - 0.5795 i & 2.0727 - 4.0857$\times 10^{-2}$ i\\
  & 0.3 & 2.2067 - 1.9686 i & 2.1067 - 0.3931 i &  1.6416 - 0.6297 i & 2.1574 - 9.2910$\times 10^{-2}$ i\\
  \hline\\
(1,\;1) %
  & 0.1 & 1.9512 - 1.5858 i & 1.4111 + 0.4430 i & 1.4277 - 0.4948 i & 1.5675 + 0.1404 i\\
  & 0.2 & 1.7679 - 1.5195 i & 0.9445 + 0.7797 i & 1.3209 - 0.4643 i & 1.2046 + 0.2445 i\\
  & 0.3 & 1.6048 - 1.4929 i & 0.5998 + 1.0384 i & 1.2273 - 0.4463 i & 0.9151 + 0.3204 i\\
  \hline\\
  (1,\;-1) %
  & 0.1 & 2.4169 - 1.8746 i & 2.7065 - 0.5962 i & 1.7067 - 0.6116 i & 2.4983 - 0.1905 i\\
  & 0.2 & 2.7349 - 2.1390 i & 3.5193 - 1.4235 i & 1.8996 - 0.7147 i & 3.0586 - 0.4532 i\\
  & 0.3 & 3.1548 - 2.5377 i & 4.4164 - 2.6136 i & 2.1522 - 0.8684 i & 3.6770 - 0.8235 i
\end{tabular}
\end{ruledtabular}
\end{table*}


In Fig.~\ref{rp1fre}, we present the real and imaginary parts of the frequency for the first boundary condition and $r_+=1$. Re($\omega$) increases with the rotation for the $m=-1$ mode, decreases with the rotation for the $m=1$ mode, and increases sightly with the rotation for the $m=0$ mode. The behaviour of Im($\omega$) is almost the opposite, since Im($\omega$) decreases with the rotation for both $m=0$ and $m=-1$ modes, but for the $m=1$ mode, it increases firstly and then starts to decrease around $a=0.33$. Comparing Fig.~\ref{rp1lam} with Fig.~\ref{rp1fre}, shows that the effect of the rotation on $\lambda$ (both real part and imaginary part) mimics closely that on $\omega$, except for  Im($\lambda$) of the $m=1$ mode, which always increases with the rotation parameter.


\subsubsection{Stationary vector clouds}
Stationary clouds are bound state solutions with real frequency of test fields around rotating background, computed at the linear level. The existence of clouds indicates nonlinear hairy BH solutions~\cite{Herdeiro:2014goa,Herdeiro:2014ima}, but the converse needs not be true~\cite{Brihaye:2014nba,Herdeiro:2015kha}. In order to find such solutions, we demand $\omega=m\Omega_H$; in other words, stationary clouds are the zero modes of superradiance. Imposing this condition leads to a constraint on the \mbox{BH} parameters: BHs are quantized in the sense that only BHs with specific parameters can support a cloud with a given set of ``quantum'' numbers. This quantization defines \textit{existence lines} in the BH parameter space. In the practical implementation of our numerical calculations, we use the same method as before, with the condition $\omega=m\Omega_H$, to look for the rotation parameter. Note that all the results presented in this subsection are for fundamental modes, characterized by $N=0$.

The vector clouds we have obtained shall be presented in a parameter space spanned by $R_+$ and $\Omega_h$, which are defined as~\cite{Henneaux:1985tv,Winstanley:2001nx,Cardoso:2013pza}
\begin{eqnarray}
R_+=\sqrt{\dfrac{r_+^2+a^2}{\Xi}}\;,\;\;\;\;\;\;\Omega_h=\Omega_H\Xi+a\;,\label{NewDef}
\end{eqnarray}
where $R_+$ approaches $r_+$ when $a$ approaches zero. The reason to use this pair of parameters, instead of  $r_+$ and $\Omega_H$, is as follows. $\Omega_H$, as defined in Eq.~\eqref{Horvel}, is the horizon angular velocity measured relatively to a rotating frame at infinity, while $\Omega_h$, defined in Eq.~\eqref{NewDef}, is the horizon angular velocity measured with respect to a non-rotating observer at infinity. The latter one is more relevant in BH thermodynamics~\cite{Gibbons:2004ai}. In the practical calculations, one can use either of them since they are simply related by Eq.~\eqref{NewDef}. As one may check, $\Omega_h$ is a monotonic function of $a$, in terms of $R_+$, but not of $r_+$. Also there is an intuitive geometric meaning for $R_+$, that is the areal horizon radius.

\begin{figure*}
\begin{center}
\begin{tabular}{c}
\hspace{-4mm}\includegraphics[clip=true,width=0.38\textwidth]{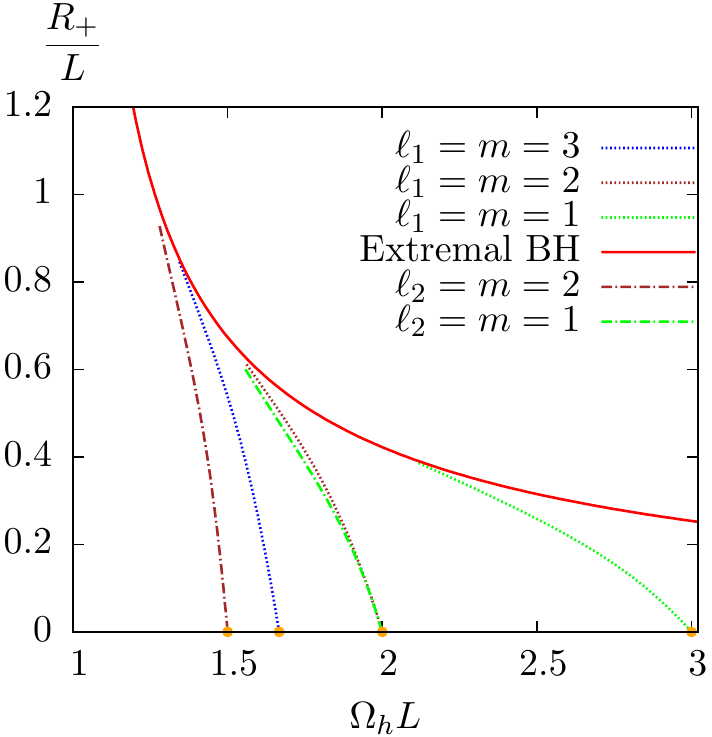}\hspace{15mm}\includegraphics[clip=true,width=0.38\textwidth]{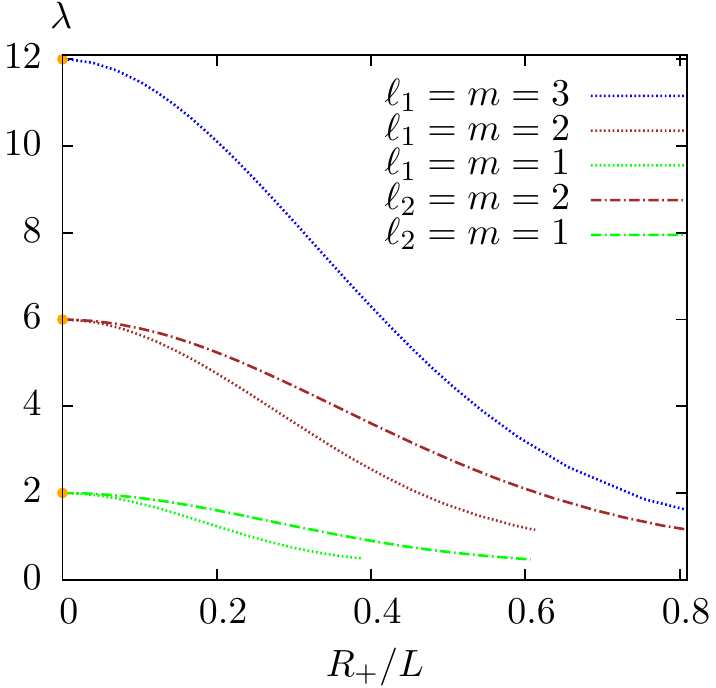}
\end{tabular}
\end{center}
\caption{\label{vec} Vector clouds (left panel) and the corresponding separation constants (right panel) in $R_+$ versus $\Omega_h$ and $\lambda$ versus $R_+$ plots, respectively. $\ell_1 (\ell_2)$ refer to the results obtained by imposing the first (second) boundary condition.}
\end{figure*}

In Fig.~\ref{vec}, the existence lines for some examples of vector clouds are displayed (left panel) together with the corresponding separation constants (right panel). In the left panel, the red solid line stands for the extremal BHs, and regular BHs only exist below this extremal line. The first three existence lines  (with $\ell=m=1,2,3$) for the first boundary condition and the first two existence lines (with $\ell=m=1,2$) for the second boundary conditions are presented by dotted and dot dashed lines, respectively. These lines start from bound state solutions (normal modes), denoted by orange dots in Fig.~\ref{vec}, of the Maxwell field on empty AdS, $i.e.$
\begin{eqnarray}
\Omega_{h,1\;|R_+=0}=1+\dfrac{2}{\ell}\;,\;\;\;\;\;\;\Omega_{h,2\;|R_+=0}=1+\dfrac{1}{\ell}\;,\label{bsAdS}
\end{eqnarray}
which are obtained by equating the superradiance condition\footnote{$\Omega_H$ is the same as $\Omega_h$ in pure AdS.}, $\omega=m\Omega_h$, to the normal mode conditions in Eqs.~\eqref{normalmode1} and~\eqref{normalmode2}, together with setting $m=\ell$, where the overtone number $N$ has been set to zero. Observe, in particular, that although the two sets of normal modes in AdS are isospectral, the existence lines for the two boundary conditions only converge as $R_+\rightarrow 0$, when taking $\ell=1$ with the first boundary condition and $\ell=2$ with the second.

An existence line with a particular $\ell=m$, separates the superradiantly stable Kerr-AdS BHs (to the left side of the existence line) and the superradiantly unstable ones (to the right side of the existence line), against that particular mode. Therefore, as one may observe from the left panel of Fig.~\ref{vec}, the stable region in the parameter space against the mode, say $\ell=m=n$, where $n$ is some integer, with the second boundary condition is also stable against all the modes with the first boundary condition from $\ell=m=1$ up to $\ell=m=n+1$. From the data in this figure, together with the relation between $R_+$ and $r_+$, eq.~(\ref{NewDef}), it can be concluded that: with the first boundary condition, BHs with $r_+\leq0.25$ are superradiantly unstable against the $\ell=m=1$ fundamental mode;  while with the second boundary condition, BHs with $r_+\leq0.34$ are superradiantly unstable against the $\ell=m=1$ fundamental mode. These observations also explain the fact that the superradiant instability only appears for the modes with the second boundary condition in Table~\ref{Table2}.

%
%

\subsubsection{Stationary scalar clouds}
\begin{figure*}
\begin{center}
\begin{tabular}{c}
\hspace{-4mm}\includegraphics[clip=true,width=0.38\textwidth]{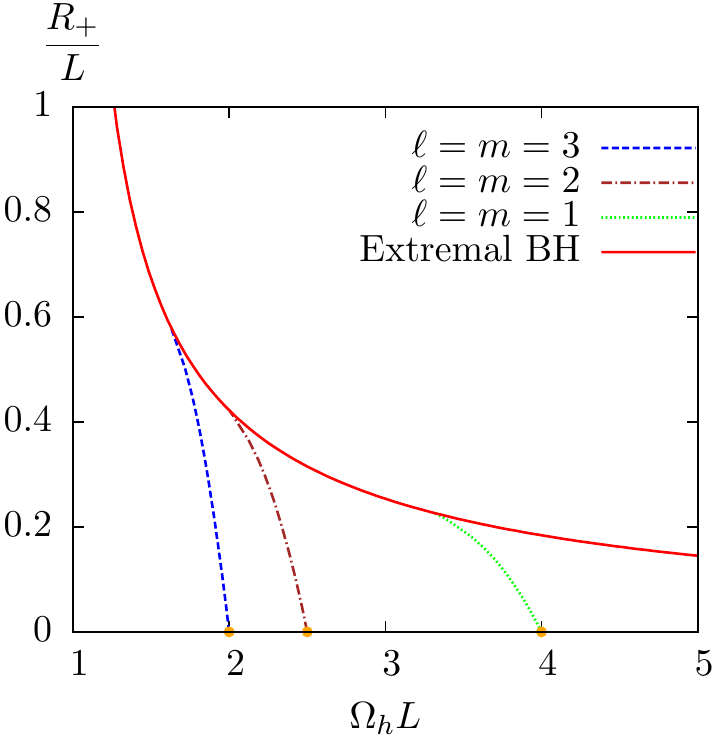}\hspace{15mm}\includegraphics[clip=true,width=0.38\textwidth]{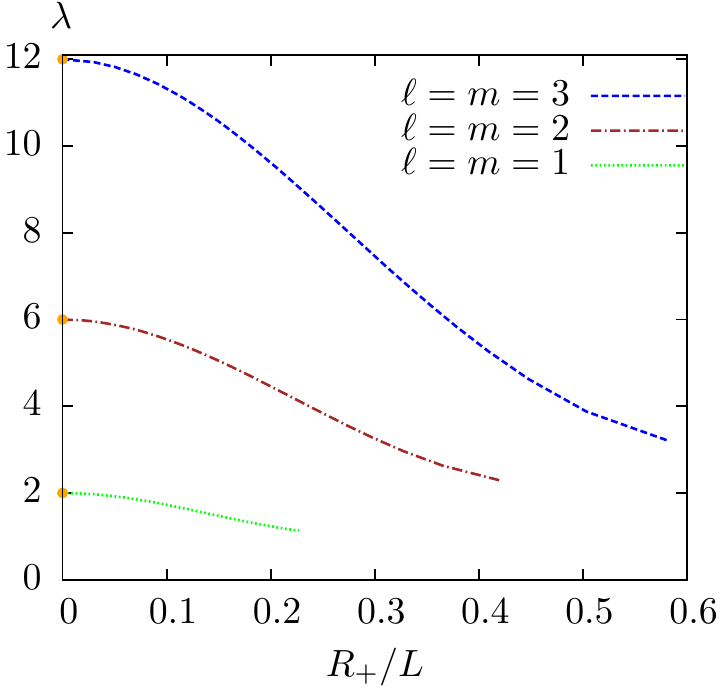}
\end{tabular}
\end{center}
\caption{\label{SCs} Scalar clouds (left panel) and the corresponding separation constants (right panel) in $R_+$ versus $\Omega_h$ and $\lambda$ versus $R_+$ plots, respectively.}
\end{figure*}

As a comparison with the Maxwell stationary clouds reported above, we have also computed stationary scalar clouds, by solving the massless Klein-Gordon equation on Kerr-AdS BHs, with vanishing energy flux boundary condition which is the same with the usual field vanishing boundary condition~\cite{Wang:2015goa}. In this case there is a single set of modes. The results for the scalar clouds are exhibited in Fig.~\ref{SCs}, in terms of the same parameters, $R_+$ and $\Omega_h$.

In the left panel of Fig.~\ref{SCs}, the red solid line stands, as before, for extremal BHs, so that regular BHs only exist below this line. The first three existence lines, corresponding to the modes with $\ell=m=1,2,3$, are described by dotted, dot dashed and dashed lines, respectively. The corresponding separation constants are also shown in the right panel of Fig.~\ref{SCs}. The orange dots in both panels stand for the normal modes and eigenvalues of the angular function in pure AdS, which are
\begin{eqnarray}
\Omega_{h,scalar}=1+\dfrac{3}{\ell}\;,\;\;\;\;\;\lambda=\ell(\ell+1)\;.
\end{eqnarray}
Again, the existence line with a particular $\ell=m$, divides the parameter space into two regions: BHs in the left region are superradiantly stable while BHs in the right region are superradiantly unstable, against that particular mode.

Comparing Figs.~\ref{vec} and~\ref{SCs} it becomes clear that the existence lines for stationary vector clouds appear to the \textit{left} of the existence line of a stationary scalar cloud with the same quantum numbers. Thus, there are BHs that are stable against the scalar mode but become unstable against the vector mode. In a sense, vector superradiance is stronger. Qualitatively, this conclusion agrees with the computation of the amplification factors for scalar and vector modes in superradiant scattering, in asymptotically flat spacetimes.


\section{Discussion and Final Remarks}
\label{discussion}

The behavior of test fields on an asymptotically AdS spacetimes depends sensitively on the boundary conditions, since such spacetimes contain a timelike boundary. The AdS boundary is often regarded as a perfectly reflecting mirror in the sense that no flux (both energy flux and angular momentum flux) can cross it. Various types of boundary conditions in asymptotically AdS spacetimes have been explored, in particular consistent with this simple requirement. But only recently this requirement was taken as the \textit{guiding principle} to impose boundary conditions~\cite{Wang:2015goa} on test fields. Following this principle, two boundary conditions are found for Maxwell fields on asymptotically AdS spacetimes, of which only one had been previously discussed on Schwarzschild-AdS background.

In this paper we have studied quasinormal modes, superradiant unstable modes and vector clouds for the Maxwell field on Kerr-AdS BHs, by imposing these two boundary conditions. To find quasinormal modes and superradiant modes, we have solved the Teukolsky equations both analytically and numerically. In the small BH and slow rotation regime, an analytical matching method was applied to exhibit how these two boundary conditions work and how they produce superradiant instabilities. A numerical method was then used to explore the parameter space where the small BH and slow rotation approximations are invalid. We find that for small BHs characterized by $r_+=0.1$, unstable superradiant modes appear with both boundary conditions. Increasing BH size, as exemplified for $r_+=0.3$, superradiant instabilities only appear with the second boundary condition, and eventually disappear for both boundary conditions, as exemplified for $r_+=1$. Our analysis also shows that superradiant instabilities for the Maxwell field may exist for (moderately) larger BH sizes, when comparing with scalar case, for which superradiant instabilities appear in the regime $r_+\leq0.16$~\cite{Uchikata:2009zz}.

To study stationary vector clouds, which can occur for massless fields in AdS, due to the box-like global structure, we have solved the Teukolsky equations at the onset of superradiant instability, $i.e.$ for $\omega=m\Omega_H$. We found that both boundary conditions can yield vector clouds, and that these clouds are bounded by the extremal BHs, as for the scalar clouds on asymptotically flat Kerr BHs~~\cite{Hod:2012px,Hod:2013zza,Herdeiro:2014goa}. This behaviour differs from that observed for gravitational perturbations, for which only one of the sets of clouds are bounded by the extremal BHs~\cite{Cardoso:2013pza}. The existence of clouds at the linear level indicates nonlinear hairy BH solutions~\cite{Herdeiro:2014goa,Herdeiro:2014ima}, so our next goal is to find the nonlinear realization of these vector clouds. There is already a well-known exact BH family within the Einstein-Maxwell-AdS system: the Kerr-Newman-AdS family. It will then be interesting to understand the interplay between this well known family and the new family of ``hairy" BHs.\footnote{A different family of static BHs, in the Einstein-Maxwell-AdS system, was reported in~\cite{Costa:2015gol} (see also~\cite{Herdeiro:2015vaa,Kichakova:2015nni}).}

\bigskip

\noindent{\bf{\em Acknowledgements.}}
We would like to thank Eugen Radu, Jo\~ao Rosa and Marco Sampaio for discussions and suggestions for this project. M.W. is  funded by FCT through the grant SFRH/BD/51648/2011. C.H. acknowledges funding from the FCT IF programme. The work in this paper is also supported by the CIDMA strategic project UID/MAT/04106/2013 and by the EU grants  NRHEP--295189-FP7-PEOPLE-2011-IRSES and H2020-MSCA-RISE-2015 Grant No. StronGrHEP-690904.

\appendix

\section{Angular momentum flux}
\label{AMF}
From the definition of the energy-momentum tensor for the Maxwell field,
\begin{equation}
T_{\mu \nu}=F_{\mu\sigma}F^\sigma_{\;\;\;\nu}+\dfrac{1}{4}g_{\mu\nu}F^2\;,\label{EMTensor}
\end{equation}
we can calculate the angular momentum flux
\begin{eqnarray}
\mathcal{J}=\int_{S^2} \sin\theta d\theta d\varphi\; r^2 \left(T^r_{\;\;\varphi,\;\uppercase\expandafter{\romannumeral1}}+T^r_{\;\;\varphi,\;\uppercase\expandafter{\romannumeral2}}\right)\;,\label{angmomf}
\end{eqnarray}
with
\begin{eqnarray}
&&T^r_{\;\;\varphi,\;\uppercase\expandafter{\romannumeral1}}=-a\sin^2\theta\;T^r_{\;\;t,\;\uppercase\expandafter{\romannumeral1}}\;,\label{relation1}\\
&&T^r_{\;\;\varphi,\;\uppercase\expandafter{\romannumeral2}}=\dfrac{i\sin\theta\sqrt{\Delta_\theta}(r^2+a^2)}{2\Xi\rho^4}\Phi_1^\ast(\Phi_2+\Delta_r\Phi_0)\nonumber\\
&&\;\;\;\;\;\;\;\;\;\;\;\;\;\;\;+c.c\;,\label{relation2}
\end{eqnarray}
where
\begin{eqnarray}
\Phi_0=\phi_0\;,\;\;\;\;\;\Phi_2=2\bar{\rho}^\ast\phi_2\;,\;\;\;\;\;\bar{\rho}=r+ia\cos\theta\;,
\end{eqnarray}
and $c.c$ stands for the complex conjugate of the proceeding terms.

From Eq.~\eqref{relation1}, and considering the vanishing energy flux boundary conditions~\cite{Wang:2015goa}, $i.e.$
\begin{equation}
\int_{S^2} \sin\theta d\theta d\varphi\; r^2 T^r_{\;\;t,\;\uppercase\expandafter{\romannumeral1}}\rightarrow0\;,
\end{equation}
asymptotically, one may conclude that there is no contributions for the angular momentum flux from the first term $T^r_{\;\;\varphi,\;\uppercase\expandafter{\romannumeral1}}$.

For the second term, from Eq.~\eqref{relation2}, we notice that $\Phi_1$ is involved so that we have to find its solution first. Since this is a lengthy derivation, we only present here the main results; the detailed proof will be shown elsewhere. The solution for $\Phi_1$ is
\begin{eqnarray}
\bar{\rho}^\ast\Phi_1=g_{+1}\bar{\mathscr{L}}_1S_{+1}-ia f_{-1}\mathscr{D}_0P_{-1}\;,
\end{eqnarray}
with
\begin{eqnarray}
&&g_{+1}=\frac{1}{B}(r\mathscr{D}_0P_{-1}-P_{-1})\;,\\
&&f_{-1}=\frac{1}{B}(\cos\theta\bar{\mathscr{L}}_1S_{+1}+\sin\theta\sqrt{\Delta_\theta}S_{+1})\;,\\
&&\bar{\mathscr{L}}_1S_{+1}=\dfrac{(2a\omega\Xi\cos\theta-\lambda)S_{+1}-BS_{-1}}{2\mathcal{Q}\sqrt{\Delta_\theta}}\;,
\end{eqnarray}
where
\begin{equation}
\mathscr{D}_0=\dfrac{\partial}{\partial r}-\dfrac{iK_r}{\Delta_r}\;,\mathcal{Q}=\dfrac{\Xi(a\omega\sin^2\theta-m)}{\sin\theta\Delta_\theta}\;,P_{-1}=BR_{-1}\;,
\end{equation}
and the constant $B$ is given by Eq.~\eqref{Beq}, $S_{+1}(\equiv S_{+1}(\theta))$ and $S_{-1}(\equiv S_{-1}(\theta))$ are spin weighted AdS spheroidal harmonics. With all of these expressions at hand, and making use of the integration properties of the spin weighted AdS spheroidal harmonics, Eq.~\eqref{relation2} becomes
\begin{eqnarray}
&&T^r_{\;\;\varphi,\;\uppercase\expandafter{\romannumeral2}}=\dfrac{i\sin\theta\sqrt{\Delta_\theta}(r^2+a^2)}{2\Xi\rho^4}(\mathcal{C}_1S_{+1}S_{-1}+\mathcal{C}_2S_{-1}^2)\;,\nonumber\\
&&\;\;\;\;\;\;\;\;\;\;\;\;\;\;\;+c.c\;,\label{Trphi2rep}
\end{eqnarray}
and the above equation should be understood under the integration. The expressions for $\mathcal{C}_1$ and $\mathcal{C}_2$ are messy in general, but they can be simplified asymptotically. The asymptotically expression for $\mathcal{C}_1$ goes as
\begin{equation}
\mathcal{C}_1\sim c_0+\mathcal{O}(1/r)\;,
\end{equation}
where $c_0$ is proportional to $T^r_{\;\;t,\;\uppercase\expandafter{\romannumeral1}}$ asymptotically, so that finally $\mathcal{C}_1\sim \mathcal{O}(1/r)$. Similar analysis can be done for $\mathcal{C}_2$ as well. The asymptotically expression for $\mathcal{C}_2$ is
\begin{equation}
\mathcal{C}_2\sim \hat{c}_0+\mathcal{O}(1/r)\;,
\end{equation}
and, as in the former case, $\hat{c}_0$ will vanish as well, after the vanishing energy flux BCs are imposed. Then from Eq.~\eqref{Trphi2rep}, we know that
\begin{eqnarray}
r^2T^r_{\;\;\varphi,\;\uppercase\expandafter{\romannumeral2}}\sim \mathcal{O}(1/r)\;,
\end{eqnarray}
asymptotically, which leads to the vanishing of the angular momentum flux, as can be seen from Eq.~\eqref{angmomf}.

\bibliographystyle{h-physrev4}
\bibliography{Maxwellbomb}


\end{document}